\documentclass[12pt]{article}
\usepackage{amsmath}
\usepackage{amssymb} % e.g. \gtrsim
\usepackage{bbm} % BlackBoeard letters
\usepackage[small]{caption2} % font of captions is different from text
\usepackage{fleqn} % equations are not centered
\usepackage{graphicx} % \including PostScript
\usepackage{mathrsfs}	% e.g. nice Lagrange-L with \mathscr{L}
\usepackage[small,loose]{subfigure}  % subfigures with (a), (b) etc., ... and own caption
\usepackage{cite} % [1-5] instead of [1,2,3,4,5]	
\advance \headheight by 3.0truept       % for 12pt mandatory...
\DeclareMathOperator{\re}{Re}
\DeclareMathOperator{\im}{Im}
\DeclareMathOperator{\tr}{tr}
\DeclareMathOperator{\Tr}{Tr}
\DeclareMathOperator{\diag}{diag}

\newcommand{\D}{\mathrm{d}}
\newcommand{\I}{\mathrm{i}}

\newcommand{\E}[1]{\ensuremath{\mathrm{E}_{#1}}} % e.g. \E{8}
\newcommand{\G}[1]{\ensuremath{\mathrm{G}_{#1}}}
\newcommand{\SO}[1]{\ensuremath{\mathrm{SO}(#1)}}
\newcommand{\SU}[1]{\ensuremath{\mathrm{SU}(#1)}}
\newcommand{\U}[1]{\ensuremath{\mathrm{U}(#1)}}
\newcommand{\Z}[1]{\ensuremath{\mathbbm{Z}_{#1}}} % Z_N ->\Z{N}

\allowdisplaybreaks

\begin{document}

\date{}
\title{\begin{flushright}
\ \\*[-80pt]
\begin{minipage}{0.3\linewidth}
\normalsize
KANAZAWA-08-04\\
KUNS-2139\\
TUM-HEP-686/08 \\*[50pt]
\end{minipage}
\end{flushright}
{\bf\huge (Non-)Abelian discrete anomalies}\\[0.8cm]}

\author{{\bf\normalsize
Takeshi Araki$^1$, Tatsuo Kobayashi$^2$, Jisuke Kubo$^1$,}\\
{\bf\normalsize
Sa\'ul Ramos-S\'anchez$^3$, Michael Ratz$^4$, Patrick~K.S.~Vaudrevange$^3$}
\\[1cm]
{\it\normalsize
${}^1$ Institute for Theoretical Physics, Kanazawa University, Kanazawa
920-1192, Japan}\\[0.2cm]
{\it\normalsize
${}^2$ Department of Physics, Kyoto University, Kyoto 606-8502, Japan}\\[0.2cm]
{\it\normalsize
${}^3$ Physikalisches Institut der Universit\"at Bonn,}\\[-0.05cm]
{\it\normalsize Nussallee 12, 53115 Bonn,
Germany}\\[0.2cm]
{\it\normalsize
${}^4$ Physik Department T30, Technische Universit\"at M\"unchen,}\\[-0.05cm]
{\it\normalsize James-Franck-Strasse, 85748 Garching, Germany} }
\thispagestyle{empty}
\maketitle

\begin{abstract}
We derive anomaly constraints for Abelian and non-Abelian discrete symmetries
using the path integral approach. We survey anomalies of discrete symmetries in
heterotic orbifolds and find a new relation between such anomalies and the
so-called `anomalous' U(1). 
\end{abstract}

\clearpage

\section{Introduction}

Symmetries play a key role in the understanding of fundamental laws of physics. 
Apart from continuous, in particular gauge, symmetries, discrete symmetries
provide a useful tool in field-theoretic model building and arise often in
top-down models.

Very much like continuous symmetries, discrete symmetries can be broken by
quantum effects, i.e.\ have an anomaly \cite{Krauss:1988zc}. If this  is the
case, one expects that the corresponding  conservation laws be violated through
non-perturbative effects. The criteria for discrete symmetries to be
non-anomalous, and thus to be exact, have been extensively studied in the
Abelian ($\Z{N}$) case \cite{Ibanez:1991hv,Banks:1991xj}. Anomaly criteria for
non-Abelian discrete symmetries have been discussed first in specific examples
\cite{Frampton:1994rk}. Here, we use the path integral approach
\cite{Fujikawa:1979ay,Fujikawa:1980eg} to derive anomaly constraints on
non-Abelian discrete symmetries. We follow the discussion of
\cite{Araki:2006mw}, and extend it such as to include gravitational anomaly
constraints. We further re-derive the conditions for Abelian discrete symmetries
to be anomaly-free, using the path integral method. This derivation allows for
an alternative, perhaps more intuitive understanding of the criteria, which does
not rely on contributions from heavy states.

We explore the issue of discrete anomalies in string compactifications,
focusing on heterotic orbifolds. The question we seek to clarify is whether
discrete anomalous symmetries can appear in string-derived models
\cite{Banks:1991xj,Dine:2004dk}. The discrete
symmetries on orbifolds reflect certain geometrical symmetries of internal
space. Since the geometrical operations, i.e.\ space group transformations, are
embedded into the gauge group, one might suspect that the discrete anomalies are
related to gauge anomalies. We find that this is indeed the case, specifically
we find that the so-called `anomalous' \U1, which occurs frequently in heterotic
orbifolds, determines the anomalies of discrete symmetries.

The paper is organized as follows. In section
\ref{sec:AnomalyFreeDiscreteSymmetries} we first re-derive anomaly constraints
for Abelian discrete symmetries and then derive the constraints for non-Abelian
discrete symmetries, using the path integral approach. In section
\ref{sec:OrbifoldAnomalies} we consider heterotic orbifolds   and identify a
geometric operation on the orbifold, which we would like to refer to as
`anomalous space group element', as the source of all discrete anomalies. 
Section \ref{sec:Summary} contains our conclusions. We also include four
appendices where we present the calculation of anomalies of the dihedral group
$D_4$ (\ref{app:D4}), $D_4$ anomalies in a concrete model from the
literature (\ref{app:GrimusLavoura}) and the anomaly coefficients in two
concrete string models (\ref{sec:KRZAnomalies} \& \ref{sec:BHLRAnomalies}).

\section{Anomaly-free discrete symmetries}
\label{sec:AnomalyFreeDiscreteSymmetries}

\subsection{A few words on symmetries}

Consider a theory described by a Lagrangean $\mathscr{L}$ with a set of 
fermions $\Psi=[\psi^{(1)},\dots,\psi^{(M)}]$,
where $\psi^{(m)}$ denotes a field transforming in the irreducible representation (irrep)
$\boldsymbol{R}^{(m)}$ of all internal symmetries.
A general transformation $\Psi~\to~U\,\Psi$
or, more explicitly,
\begin{equation}\label{eq:Trafo2}
 \left[\begin{array}{c}
 \psi^{(1)}\\ \vdots\\ \psi^{(M)}\end{array}\right]
 ~\to~
 \left(\begin{array}{ccc}
 U^{(1)} &  & 0\\
 & \ddots & \\
 0 & & U^{(M)} 
 \end{array}\right)\,
 \left[\begin{array}{c}
 \psi^{(1)}\\ \vdots\\ \psi^{(M)}\end{array}\right]\;,
\end{equation}
which leaves $\mathscr{L}$ invariant (up to a total derivative) denotes a
classical symmetry.
By Noether's theorem, \emph{continuous} symmetries imply, at the classical
level, conserved currents, % $j^\mu$, 
$D_\mu j^\mu~=~0$.
For instance, in the case of an Abelian continuous symmetry one can define the 
charge $Q~=~\int\!\D^3 x\,j^0$
which satisfies the conservation law $\frac{\D}{\D t}Q~=~0$.

In the case of a discrete symmetry, the situation is similar. Consider, for simplicity, an Abelian
discrete symmetry, i.e.\ \Z{N}. Under this symmetry, the fermions of the theory transform as
\begin{equation}\label{eq:ZN}
 \psi^{(m)}~\to~\mathrm{e}^{2\pi\,\I\,q^{(m)}/N}\,\psi^{(m)}\;,
\end{equation}
where (by convention) the discrete charges $q^{(m)}$ are integer and only
defined modulo $N$. If \eqref{eq:ZN} is a symmetry of $\mathscr{L}$, 
the corresponding charge $q^{(m)}$
is conserved modulo $N$.\footnote{A familiar example for such a
conservation law is due to $R$-parity, which implies that superpartners can only be
produced in pairs.}

\subsection{Basics of anomalies}

\begin{figure}[t!]
 \centerline{%
 	\subfigure[$\U1-G-G$.\label{fig:AnomalyU1-G-G}]{\includegraphics{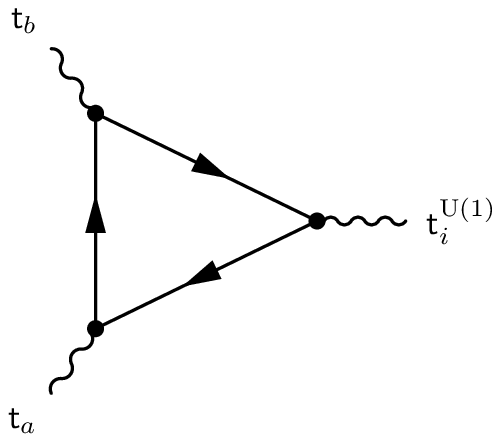}}
	\quad
	\subfigure[$D-G-G$.\label{fig:AnomalyD-G-G}]{\includegraphics{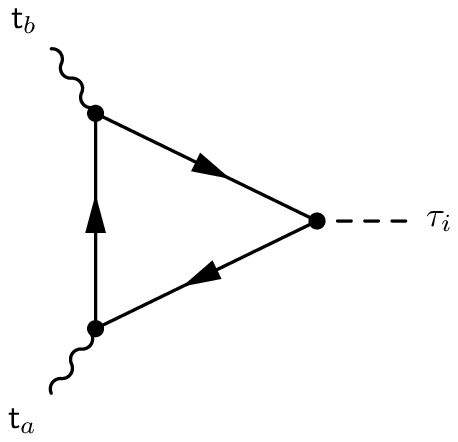}}
 }
\caption{Triangle diagrams.} 
\label{fig:TriangleDiagrams} 
\end{figure}

Classical chiral symmetries  can be broken by quantum effects, i.e.\ have an
anomaly. Specifically, consider a chiral transformation
\begin{equation}\label{eq:ChiralTransformation}
 \Psi(x) ~ \to ~ \Psi'(x)~=~\exp\big(\I\,\alpha\,P_\mathrm{L}\big)\,\Psi(x)\;,
\end{equation}
where $\alpha=\alpha^A\mathsf{T}_A$ with $\mathsf{T}_A$ denoting the generators
of the transformation, and $P_\mathrm{L}$ is the left-chiral projector. It is
well known that at the quantum level the classically conserved current
$j^\mu(x)$ is not necessarily conserved any more, that is (cf.\ e.g.\
\cite{Bertlmann:1996xk})
\begin{equation}\label{eq:Anomaly}
 \langle D_\mu  j^\mu(x)\rangle~=~\mathcal{A}(x;\alpha)~\ne~0\;.
\end{equation}
The anomaly $\mathcal{A}(x;\alpha)$  can be derived using Fujikawa's method,
i.e.\ by calculating the transformation of the path integral measure
\cite{Fujikawa:1979ay,Fujikawa:1980eg},     which in our case reads
\begin{equation}
 \mathcal{D}\Psi\,\mathcal{D}\overline{\Psi}
 ~\to~J(\alpha)\,
 \mathcal{D}\Psi\,\mathcal{D}\overline{\Psi}\;,
\end{equation}
where the Jacobian of the transformation is given by
\begin{equation}\label{eq:Jacobian}
 J(\alpha)
 ~=~\exp\left\{\I\,\int\!\D^4x\,\mathcal{A}(x;\alpha)\right\}\;.
\end{equation}
The anomaly function $\mathcal{A}$ decomposes into a gauge and a
gravitational part
\cite{AlvarezGaume:1983ig,AlvarezGaume:1984dr,Fujikawa:1986hk},
\begin{equation}\label{eq:AnomalyFunctionA}
 \mathcal{A}~=~
  \mathcal{A}_\mathrm{gauge}+\mathcal{A}_\mathrm{grav}
\;.
\end{equation}
The gauge part $\mathcal{A}_\mathrm{gauge}$ corresponds to the triangle diagram 
$\alpha$--gauge--gauge (figure~\ref{fig:TriangleDiagrams}). 
This anomaly is given by\footnote{Note that there is a factor 1/2 discrepancy to
Fujikawa's result~\cite{Fujikawa:1979ay,Fujikawa:1980eg} because we are
considering only fermions of one chirality (cf.\ e.g.\ \cite{Fujikawa:2004cx}
and \cite[p.~271]{Bertlmann:1996xk}).} 
\begin{equation}
 \mathcal{A}_\mathrm{gauge}(x;\alpha)~=~\frac{1}{32\,\pi^2}
 \Tr\left[\alpha\,\mathcal{F}^{\mu\nu}(x)\,\widetilde{\mathcal{F}}_{\mu\nu}(x)\right]\;.
\label{eq:Agauge}
\end{equation}
Here $\mathcal{F}_{\mu\nu}=[D_\mu,D_\nu]$ is the field strength of the gauge symmetry, such that
$\mathcal{F}_{\mu\nu}=g\,(\partial_\mu A_\nu-\partial_\nu A_\mu)$ for a \U1 symmetry, and
$\widetilde{\mathcal{F}}^{\mu\nu}=\varepsilon^{\mu\nu\rho\sigma}\mathcal{F}_{\rho\sigma}$
denotes its dual. 
The trace `Tr' runs over all internal indices. 

Analogously, the gravitational part $\mathcal{A}_\mathrm{grav}$ is the mixed
$\alpha$--gravity--gravity anomaly. It is known that it takes the form
\cite{AlvarezGaume:1983ig,AlvarezGaume:1984dr,Fujikawa:1986hk}
\begin{equation}
 \mathcal{A}_\mathrm{grav}
 ~=~
 -\mathcal{A}_\mathrm{grav}^\mathrm{Weyl\:fermion} 
  \sum_{m}
   \tr \left[\alpha(\boldsymbol{R}^{(m)})\right]\;, 
\label{eq:Agrav}
\end{equation}
where the summation runs over the (spin-1/2) fermions in the representations
$\boldsymbol{R}^{(m)}$. The subscript `$m$' indicates that
each representation $\boldsymbol{R}^{(m)}$ appears only once in the sum.
$\alpha(\boldsymbol{R}^{(m)})$ denotes
$\alpha_A\,\mathsf{T}^A(\boldsymbol{R}^{(m)})$ in the representation
$\boldsymbol{R}^{(m)}$, and might therefore be thought of as a $\dim
\boldsymbol{R}^{(m)}\times\dim \boldsymbol{R}^{(m)}$ matrix such that `$\tr$' is
the standard (matrix) trace. 
The contribution  of a single Weyl fermion to the
gravitational anomaly is given by \cite{AlvarezGaume:1983ig,AlvarezGaume:1984dr,Fujikawa:1986hk}  
\begin{equation}
 \mathcal{A}_\mathrm{grav}^\mathrm{Weyl\:fermion}  ~=~
 \frac{1}{384 \pi^2}\,\frac{1}{2}\,
 \varepsilon^{\mu\nu\rho\sigma}\, R_{\mu\nu}{}^{\lambda\gamma}\,
 R_{\rho\sigma\lambda\gamma}\;.
\end{equation}

To evaluate the anomaly~\eqref{eq:AnomalyFunctionA}, we split the set of all
generators $\mathsf{T}_A$ into generators of continuous symmetries
$\mathsf{t}_a$ and those of discrete symmetries $\tau_i$.
Therefore, we shall discuss separately the two cases:
\renewcommand{\labelenumi}{(\roman{enumi})}
\begin{enumerate}
 \item anomalies of continuous symmetries with $\alpha=\alpha^a \mathsf{t}_a$; 
 \item anomalies of discrete symmetries with $\alpha=\alpha^i\tau_i$.
\end{enumerate}
Note that we (implicitly) assume that all symmetries are gauged.

For the evaluation of the anomalies, it is useful to recall the
powerful index theorems~\cite{AlvarezGaume:1983ig,AlvarezGaume:1984dr}, which
imply
\begin{subequations}\label{eq:indexTheorems}
\begin{eqnarray}
 \int\!\D^4 x\, \frac{1}{32\pi^2}\,\varepsilon^{\mu\nu\rho\sigma}\,
 	F_{\mu\nu}^a\,F_{\rho\sigma}^b\,
	\tr\left[\mathsf{t}_a\,\mathsf{t}_b\right]
 & \in & \mathbbm{Z}
 \;, \label{eq:index1}\\*
 \frac{1}{2}\,\int\!\D^4 x\, \frac{1}{384 \pi^2}\,\frac{1}{2}\,
 \varepsilon^{\mu\nu\rho\sigma}\, R_{\mu\nu}{}^{\lambda\gamma}\,
 R_{\rho\sigma\lambda\gamma}
 & \in & \mathbbm{Z}
 \;,\label{eq:index2}
\end{eqnarray}
\end{subequations}
where $\mathsf{t}_a$ are in the fundamental representation of a particular gauge factor $G$.
Note that in our conventions $\tr[\mathsf{t}_a\,\mathsf{t}_b]=\frac{1}{2}\delta_{ab}$.
The factor $\frac12$ in eq.~\eqref{eq:index2} follows from Rohlin's theorem~\cite{Rohlin:1959}, as
discussed in~\cite{Csaki:1997aw}.

\subsection{Anomaly constraints for continuous symmetries}

We start by reviewing the  anomaly constraints for the continuous symmetries.
They arise from demanding that $\mathcal{A}(x;\alpha^a\mathsf{t}_a)$ vanish for
arbitrary $\alpha^a$ in order for the Jacobian  $J(\alpha)$ to be trivial. 
Consider first the mixed $\U1-G-G$ anomaly, where 
$G$ is a non-Abelian gauge factor with generators $\mathsf{t}_a$.
Representations under $G$ are denoted by $\boldsymbol{r}^{(f)}$.
This anomaly can be related to diagram~\ref{fig:AnomalyU1-G-G}.
From equation~\eqref{eq:Agauge} and the index theorem~\eqref{eq:index1}, one finds that
it only vanishes if
\begin{equation}\label{eq:A[U(1)-G-G]}
 A_\mathrm{\U1-G-G}~\equiv~
  \sum_{\boldsymbol{r}^{(f)}} q^{(f)}\,\ell(\boldsymbol{r}^{(f)})~=~0\;.
\end{equation}
In analogy to equation~\eqref{eq:Agrav}, `$\sum_{\boldsymbol{r}^{(f)}}$' means
that each representation $\boldsymbol{r}^{(f)}$ is only summed once.\footnote{
Of course, the dimensions of representations w.r.t.\ further symmetry factors 
have to be taken into account.}  
$q^{(f)}$ denote the respective \U1 charges. 
The Dynkin indices $\ell(\boldsymbol{r}^{(f)})$ are defined by
\begin{equation}
 \ell(\boldsymbol{r}^{(f)})\,\delta_{ab}
 ~=~\tr\left[\mathsf{t}_a\left(\boldsymbol{r}^{(f)}\right)\,
 	\mathsf{t}_b\left(\boldsymbol{r}^{(f)}\right)\right]
 \;.
\end{equation}
Our conventions are such that $\ell(\boldsymbol{M})=1/2$ for \SU{M} and
$\ell(\boldsymbol{M})=1$ for \SO{M}.
Consider next the $\U1-\mathrm{grav}-\mathrm{grav}$ anomaly, equation~\eqref{eq:Agrav}. 
From the index theorem~\eqref{eq:index2}, 
it vanishes if 
\begin{equation}
 A_\mathrm{\U1-\mathrm{grav}-\mathrm{grav}}
 ~\equiv~ 
 \sum_{f} q^{(f)}
 ~=~
 \sum_{m}
 q^{(m)}\,\dim(\boldsymbol{R}^{(m)})
 ~=~0\;.
\end{equation}
The sum `$\sum_f$' indicates a plain summation over all fermions.

In summary, we see that the continuous symmetries are non-anomalous if and only if
the Jacobian~\eqref{eq:Jacobian} is trivial for arbitrary $\alpha$.

\subsection{(Re-)Derivation of anomaly constraints for $\boldsymbol{\mathbbm{Z}_N}$ symmetries}

Now consider a discrete symmetry, i.e.\ $\alpha=\alpha^i\tau_i$ where, by convention,
$\alpha^i$ takes only the discrete values $2\pi/N^i$ and the eigenvalues of $\tau_i$ are integer. 
Like before, we demand that $J(\alpha)$ be
trivial. It is now important to note that the Jacobian can also be trivial for
non-zero arguments of the exponential. 
Let us specify the conditions for this to happen.
Consider first the Abelian case, i.e.\ a \Z{N} symmetry with $\alpha=2\pi\,\tau/N$.
From the gauge and gravitational parts of the anomaly
function, equations~\eqref{eq:Agauge} and~\eqref{eq:Agrav}, and the index 
theorems~\eqref{eq:indexTheorems}, we see
that the Jacobian is trivial if
\begin{subequations}
\begin{eqnarray}
 A_{\Z{N}-G-G} & = & \frac{1}{N}\sum_{\boldsymbol{r}^{(f)}} q^{(f)} \,
 \big(2\,\ell(\boldsymbol{r}^{(f)})\big) ~\in~\mathbbm{Z}
 \;,\label{eq:A_Z_N-G-G}\\
 A_{\Z{N}-\mathrm{grav}-\mathrm{grav}} &= & 
 \frac{2}{N}
 \sum_{m} q^{(m)} \,
 \dim\boldsymbol{R}^{(m)}
 ~\in~\mathbbm{Z}\; .\label{eq:A_Z_N-grav-grav}
\end{eqnarray}
\end{subequations}
The factor 2 in front of the Dynkin index in \eqref{eq:A_Z_N-G-G} is due to our
conventions ($\tr[\mathsf{t}_a\,\mathsf{t}_b]=\frac{1}{2}\delta_{ab}$).
This means that the constraints for a \Z{N} symmetry to be anomaly-free are
\begin{subequations}\label{eq:ZNconditions}
\begin{eqnarray}
 \Z{N}-G-G & : &
 \sum_{\boldsymbol{r}^{(f)}} q^{(f)} \,
 \ell(\boldsymbol{r}^{(f)}) ~=~ 0\mod N/2\;,\label{eq:condition-gauge}\\
 \Z{N}-\mathrm{grav}-\mathrm{grav} & : &
 \sum_{m}
 q^{(m)} \,\dim\boldsymbol{R}^{(m)} ~=~0\mod N/2\;.\label{eq:condition-grav}
\end{eqnarray}
\end{subequations}
If $N$ is odd, we can always make the \Z{N} charges even by shifting them by
integer multiples of $N$. This explains why the sums in
\eqref{eq:condition-gauge} and \eqref{eq:condition-grav} can always be made
integer. Hence in the case of an odd $N$ one can replace $N/2$ by $N$ after
a suitable shift of the charges. The
constraints \eqref{eq:condition-gauge} and \eqref{eq:condition-grav} coincide
with the ones of the literature
\cite{Ibanez:1991hv,Banks:1991xj,Ibanez:1991pr,Ibanez:1992ji,Csaki:1997aw,Babu:2002tx,Dreiner:2005rd}.
We would like to emphasize that, in our derivation, we did not invoke the
contributions from heavy Majorana fermions.\footnote{The mod~$N/2$ condition
for even $N$ has been justified as follows \cite{Ibanez:1991hv}: one can always
introduce Majorana fermions $\psi$ with $\Z{N}$ charges $N/2$. Their contribution to
the sum is $N/2$, on the other hand the Majorana mass term $m\,\psi\,\psi$ is
allowed by the discrete symmetry. Since $m$ can be arbitrarily large, $\psi$
can be `removed from the theory'. While the argument leads to the correct
result, one might nevertheless wonder if the anomaly conditions change if one
considers more constrained settings (such as string-derived theories) where
extra degrees of freedom cannot be introduced at will. Our derivation shows that
the anomaly conditions remain unchanged.} Rather, the anomaly constraints
(including the condition mod$\,N/2$) are a consequence of the index theorems and
follow from demanding that the Jacobian be trivial. We also note that in our
approach one immediately sees that there are no cubic anomaly constraints for
discrete symmetries, which is in agreement with \cite{Banks:1991xj}.

\subsection{Anomalies of non-Abelian discrete symmetries}
\label{sec:NonAbelianDiscretAnomalies}

We now turn to non-Abelian discrete symmetries $D$. Consider a specific
transformation $U$. Since we are considering a discrete symmetry, there is a
positive integer $N$ such that $U^N=\mathbbm{1}$, i.e.\ $U$ is generating a
\Z{N} symmetry; we take $N$ to be the smallest such integer. 
Denote the (discrete) representations of $D$ by $\boldsymbol{d}^{(f)}$. Moreover, 
an element $U\in D$ in a representation $\boldsymbol{d}^{(f)}$ is given by
$U(\boldsymbol{d}^{(f)})=\mathrm{e}^{\I\,\alpha(\boldsymbol{d}^{(f)})}$ with
$\alpha(\boldsymbol{d}^{(f)})=2\pi\,\tau(\boldsymbol{d}^{(f)})/N$
and $\tau(\boldsymbol{d}^{(f)})$ having integer eigenvalues.
In the evaluation of the anomaly functions, equations~\eqref{eq:Agauge} and~\eqref{eq:Agrav}, 
we note that $\tr[\tau(\boldsymbol{d}^{(f)})]$ takes the role of
the \Z{N} charge. This charge, denoted by $\delta^{(f)}$,
can be expressed in terms of the group elements $U(\boldsymbol{d}^{(f)})$ as 
(cf.\ \cite{Araki:2006mw})
\begin{equation}\label{eq:trtaui}
 \delta^{(f)}~\equiv~\tr\left[\tau(\boldsymbol{d}^{(f)})\right] 
 ~=~N\,\frac{\ln \det U(\boldsymbol{d}^{(f)})}{2\pi\,\I}\;.
\end{equation}
As usual, the \Z{N} charges $\delta^{(f)}$ are defined modulo $N$ only (such
that they can consistently be expressed through the multi-valued logarithm).

From the index theorems~\eqref{eq:indexTheorems}, we find that demanding that
the Jacobian be trivial amounts to requiring 
\begin{subequations}\label{eq:NonAbelianDiscreteAnomalyConstraints1}
\begin{eqnarray}
 \sum_{(\boldsymbol{r}^{(f)},\boldsymbol{d}^{(f)})} 
 \delta^{(f)}\cdot \ell(\boldsymbol{r}^{(f)})
 & \stackrel{!}{=} &
 0\mod \frac{N}{2}\;,\label{eq:NAcondition-gauge2}\\
 \sum_{\boldsymbol{d}^{(f)}}\delta^{(f)}
 & \stackrel{!}{=} &
 0\mod \frac{N}{2}\;,
 \label{eq:NAcondition-grav2}
\end{eqnarray}
\end{subequations}
where the sum `$\sum_{(\boldsymbol{r}^{(f)},\boldsymbol{d}^{(f)})} $'
indicates that only over representations is summed which are non-trivial w.r.t.\
both $G$ and $D$; the symbol `$\sum_{\boldsymbol{d}^{(f)}}$' in
\eqref{eq:NAcondition-grav2} means that the sum extends over all non-trivial
representations $\boldsymbol{d}^{(f)}$.

These constraints have to be fulfilled for each discrete transformation $U$
separately. However, elements with $\det U =1$ do not lead to anomalies, 
cf.\ equation~\eqref{eq:trtaui}.

Non-Abelian discrete groups $D$ have more than one element. Assume that
we have verified that the constraints
\eqref{eq:NonAbelianDiscreteAnomalyConstraints1} are fulfilled for $U,~U'\in D$.
It is then obvious that this implies that for both elements $U''=U\cdot U'$ and
$U'''=U'\cdot U$ equations \eqref{eq:NonAbelianDiscreteAnomalyConstraints1} hold
as well. This means that in practice one only has to check anomaly constraints
for the generators of $D$.
 In
appendix \ref{app:D4}, we discuss $D_4$ anomalies as an example for non-Abelian
discrete anomalies. In appendix \ref{app:GrimusLavoura} we present a sample calculation.

\subsection{Summary of anomaly constraints}

The anomaly constraints for discrete symmetries can be summarized as follows:
\begin{enumerate}
 \item anomalies of \Z{N} symmetries
\begin{subequations}\label{eq:ZNAnomalyConstraints}
\begin{eqnarray}
 \Z{N}-G-G & : & 
 \sum_{\boldsymbol{r}^{(f)}} q^{(f)}\cdot \ell(\boldsymbol{r}^{(f)})   
 ~ \stackrel{!}{=} ~
 0\mod \frac{N}{2}\;,\\
  \Z{N}-\mathrm{grav}-\mathrm{grav} & : & 
 \sum_{m}q^{(m)}\cdot \dim\boldsymbol{R}^{(m)} 
 ~ \stackrel{!}{=} ~ 0\mod \frac{N}{2}\;.
\end{eqnarray}
\end{subequations}

 \item anomalies of non-Abelian discrete symmetries $D$: one has to verify that
 for all generators of $D$ the following two equations hold
\begin{subequations}\label{eq:NonAbelianAnomalyConstraints}
\begin{eqnarray}
 D-G-G & : & 
 \sum_{(\boldsymbol{r}^{(f)},\boldsymbol{d}^{(f)})} 
 \delta^{(f)}\cdot  \ell(\boldsymbol{r}^{(f)})
 ~ \stackrel{!}{=} ~ 0\mod \frac{N}{2}\;,\\
  D-\mathrm{grav}-\mathrm{grav} & : & 
 \sum_{\boldsymbol{d}^{(f)}} \delta^{(f)}
 ~ \stackrel{!}{=} ~
 0\mod \frac{N}{2}\;.
\end{eqnarray}
\end{subequations}
Here, the sum $\sum_{\boldsymbol{d}^{(f)}}$ extends over all non-trivial
representations of $D$, $\delta^{(f)}$ is defined in equation \eqref{eq:trtaui},
and $N$ denotes the order of the generator.
\end{enumerate}

\subsection{Consequences of discrete anomalies}
\label{sec:DiscreteAnomalies}

Now we turn to study the implications of an anomalous discrete symmetry.
One might envisage several scenarios in which such a symmetry appears. In what
follows, we focus on a particular one: we start with a so-called `anomalous' \U1
and break it to a discrete subgroup. Later, in
section~\ref{sec:OrbifoldAnomalies}, where we investigate string-derived models,
we will attempt to realize different situations.

In a fundamental theory, anomalies of a continuous symmetry are not acceptable.
However, there is the well-understood situation in which a \U1 factor appears
`anomalous', i.e.\ the usual anomaly conditions seem not to be satisfied. This is
the case when the anomaly is canceled by the Green-Schwarz (GS) mechanism
\cite{Green:1984sg}. To discuss this scenario, consider a supersymmetric gauge
theory. Under the (`anomalous') $\U1_\mathrm{anom}$ transformation, the chiral
superfields  $\Phi^{(f)}$ containing the chiral fermions $\psi^{(f)}$ and the
vector superfield $V$ transform as
\begin{equation}
 \Phi^{(f)}~\to~\mathrm{e}^{-\I\,q^{(f)}\,\Lambda}\,\Phi^{(f)}\;,\quad
 V~\to~V+\I\,\left(\Lambda-\overline{\Lambda}\right)\;.
\end{equation}
The anomaly is canceled by the transformation of the dilaton $S$ (or possibly a
different chiral field), which gets shifted under the $\U1_\mathrm{anom}$
transformation as
\begin{equation}\label{eq:DilatonTransformation}
S~\to~S+\frac{\I}{2}\delta_\mathrm{GS}\,\Lambda\;,
\end{equation}
where $\delta_\mathrm{GS}$ is proportional to the trace of the generator of
$\U1_\mathrm{anom}$, $\tr\mathsf{t}_\mathrm{anom}$, (see below).
The tree-level K\"ahler potential for the dilaton is 
\begin{equation}\label{eq:Kdilaton-0}
K_\mathrm{dilaton}\left(S+\overline{S}\right)~=~-\ln 
\left(S+\overline{S}\right) \;.
\end{equation}
As usual, the kinetic terms for the scalar components of $S$ arise from the
corresponding $D$-term, 
$[K_\mathrm{dilaton}\left(S+\overline{S}\right)]_D$, i.e. 
\begin{equation}
\label{eq:kin-dilaton-axion}
\frac{1}{4s^2}\left ( \partial^\mu s\, \partial_\mu s + 
\partial^\mu a \,\partial_\mu a \right)\;,
\end{equation}
where $s=\re S$ and $a=\im S$.
Consider now the axionic shift \eqref{eq:DilatonTransformation},  
\begin{equation}\label{eq:axion-shift}
 a~\rightarrow~a + \theta/2\;.
\end{equation}
The kinetic term \eqref{eq:kin-dilaton-axion} is invariant 
under this shift when $\theta$ is constant.
However, as the parameter $\theta$ depends on $x$ for $\U1_\mathrm{anom}$
transformations, the kinetic term~\eqref{eq:kin-dilaton-axion} is not 
invariant under $\U1_\mathrm{anom}$.
To make it invariant, we have to introduce the terms,  $A^\mu\partial_\mu a$ and
$A^\mu A_\mu$, in the  St\"uckelberg form.
That implies the $\U1_\mathrm{anom}$-invariant 
K\"ahler potential for the dilaton is
\begin{equation}\label{eq:Kdilaton}
 K_\mathrm{dilaton}\left(S+\overline{S}-\frac{\delta_\mathrm{GS}}{2}V\right)\;,
\end{equation}
which also includes the $s$-dependent Fayet-Iliopoulos (FI) D-term.

It is convenient to define a normalized $\U1_\mathrm{anom}$ generator
$\widetilde{\mathsf{t}}_\mathrm{anom}$,
whose charges $\widehat{q}_\mathrm{anom}$ fulfill the consistency conditions (cf.\
\cite{Kobayashi:1996pb}) 
\begin{equation}\label{eq:deltaGS1}
 \frac{1}{3}\sum_{f} \left(\widehat{q}_\mathrm{anom}^{(f)}\right)^3
 ~=~
 \frac{1}{24}\sum_{f} \widehat{q}_\mathrm{anom}^{(f)}
 ~=:~8\pi^2\,\delta_\mathrm{GS}\;.
\end{equation}
Our conventions are such that $\delta_\mathrm{GS}$ is positive.
(As before, $\sum_f$ means plain summation.)
The mixed $\U1_\mathrm{anom}-G-G$ anomaly coefficients, as defined in
\eqref{eq:A[U(1)-G-G]}, have to satisfy the consistency conditions
\begin{equation}
 \frac{1}{k}A_{\U1_\mathrm{anom}-G-G}
 ~=~
 8\pi^2\,\delta_\mathrm{GS}\;,
\end{equation}
where $k$ denotes the Ka\v{c}-Moody level of $G$. For the Green-Schwarz mechanism to
work, this relation has to hold for all gauge group factors.

The first question is whether $\U1_\mathrm{anom}$ can be used to forbid
couplings. To answer this question, consider a product of fields,
$\Phi^{(1)}\cdots\Phi^{(n)}$, with $\sum_i q_i<0$. In the case of a usual \U1 symmetry,
$\Phi^{(1)}\cdots\Phi^{(n)}$ cannot denote an allowed coupling. However, in the case
of $\U1_\mathrm{anom}$, this conclusion does not apply; instead one finds that
the non-perturbative coupling
\begin{equation}\label{eq:NPterms}
 \mathrm{e}^{-p\,S/\delta_\mathrm{GS}}\,\Phi^{(1)}\cdots\Phi^{(n)}
\end{equation}
with an appropriate $p$ can be induced (cf.\ 
\cite{Banks:1991xj,Banks:1995ii,ArkaniHamed:1998nu}).
In other words, the field $\Sigma=\mathrm{e}^{-p\,S/\delta_\mathrm{GS}}$ transforms under the \U1 with a
charge that is opposite to $\tr q_\mathrm{anom}$.
That means that $\U1_\mathrm{anom}$ does not forbid products of fields with
$\sum_i q_i<0$. 

What can one say about products of fields $\Phi^{(1)}\cdots\Phi^{(n)}$ with
$\sum_i q_i>0$? Here the answer is that an anomalous \U1 implies the existence
of a FI  $D$-term (cf. equation~\eqref{eq:Kdilaton}). To obtain a supersymmetric vacuum, the FI
term has to be canceled. That is, certain fields with net negative anomalous
charge have to attain a VEV in the vacuum. Multiplication of
$\Phi^{(1)}\cdots\Phi^{(n)}$ by such fields can lead to allowed couplings, hence
in supersymmetric vacua couplings of the type $\Phi^{(1)}\cdots\Phi^{(n)}$ will
generically be allowed.

Given these considerations, it is also clear what happens if one breaks
$\U1_\mathrm{anom}$ to a discrete, anomalous subgroup. Since $\U1_\mathrm{anom}$
is violated by terms of the form \eqref{eq:NPterms}, also the discrete subgroup
is expected not to be exact.\footnote{A special situation arises if
$\U1_\mathrm{anom}$ gets broken to a \Z{N} subgroup which, however, is
non-anomalous by the criteria \eqref{eq:ZNconditions}. Here, either the terms
\eqref{eq:NPterms} appear nevertheless, or there is a subclass of terms, which
are forbidden by the non-anomalous \Z{N}, and where the coefficient happens to
be zero. That is, if the second possibility is true, non-perturbative effects
break $\U1_\mathrm{anom}$ to an non-anomalous \Z{N} subgroup. To find out which
situation is realized would be, by itself, an interesting question, which is,
however, beyond the scope of this study.}

An anomaly of an Abelian discrete symmetry does not necessarily signal an
inconsistency of the model. Symmetries might just be accidental or approximate,
and, therefore, need not to be gauged. Further, if the anomalies are universal,
they can be canceled by a Green-Schwarz mechanism. In practice, this means that
they are broken by the VEVs of certain fields; in addition there are
non-perturbatively induced terms with hierarchically small
coefficients, as in \eqref{eq:NPterms}. These small corrections might turn out
to be a virtue rather than a problem in concrete models.

\subsection{A comment on the `SUSY zero mechanism'}

We conclude this section by commenting on supersymmetric texture zeros
\cite{Leurer:1993gy,Binetruy:1996xk}, 
which go sometimes also under the term `SUSY zero mechanism' .
It is stated that, due to holomorphicity of the superpotential, an
anomalous \U1 symmetry can enforce absence of certain couplings even though the
symmetry is broken in supersymmetric vacua, where the FI $D$-term
is canceled. Let us briefly review the argument: cancellation of the FI term
requires certain field with certain, say negative, sign of `anomalous' charge to
attain a vacuum expectation value (VEV). Now one might envisage a situation in
which only fields with non-positive charges get a VEV. Consider then a
combination of some other fields, $\Phi^{(1)}\cdots\Phi^{(n)}$, which has total negative
anomalous charge. To be neutral w.r.t.\ the $\U1_\mathrm{anom}$ symmetry, this
combination needs to be multiplied by fields with positive $\U1_\mathrm{anom}$
charge. However, so the argument goes, those fields do not attain VEVs, and
hence $\Phi^{(1)}\cdots\Phi^{(n)}$ cannot denote an allowed coupling. That is, couplings
of the type $\Phi^{(1)}\cdots\Phi^{(n)}$ appear to be absent. On the other hand, in many
applications of the `SUSY zero mechanism' it is not possible to specify a
symmetry that forbids those couplings. 

With what we have discussed above,  we are able to resolve the puzzle:
$\Sigma=\mathrm{e}^{-p\,S/\delta_\mathrm{GS}}$ carries positive charge and hence couplings of the form
$\Sigma\,\Phi^{(1)}\cdots\Phi^{(n)}$ can arise. The induced effective coupling is
suppressed (so that, as far as textures are concerned, a `zero' can be a good
approximation), however, in contrast to what is often assumed, in general it is not
related to the scale of supersymmetry breakdown.

\section{Anomalies in heterotic orbifold models}
\label{sec:OrbifoldAnomalies}

An interesting question is whether discrete anomalies occur in top-down
constructions, in particular in string compactifications
\cite{Banks:1991xj,Dine:2004dk}. Since string theory is believed to be UV
complete, one would expect that there are no (uncanceled) anomalies in this
framework. While this has been extensively checked for continuous gauge
symmetries, the case of discrete symmetries is somewhat more subtle.
Construction of string models exhibiting discrete anomalies would lead to a
playground in which the `quantum gravity effects', which are commonly believed
to spoil the discrete conservation laws, can be specified in somewhat more
detail than usual.

Specifically, we study anomalies of discrete symmetries in heterotic orbifold
models. In our presentation, we mainly focus on the \Z6-II orbifold, yet in our
computations we also considered different orbifolds, so that our results are more
generally valid. We start with a very brief review on orbifolds, summarize the
essentials of (discrete) string selection rules, continue by relating the
so-called `anomalous \U1' to a discrete transformation in compact space, which
we refer to as the `anomalous space group element $g^\mathrm{anom}$', and
conclude by relating anomalies in discrete symmetries to the anomaly in the
discrete transformation $g^\mathrm{anom}$.

\subsection{Orbifold basics}

A heterotic orbifold emerges by dividing a six-dimensional torus $\mathbbm{T}^6$
by one of its symmetries $\theta$ \cite{Dixon:1985jw,Dixon:1986jc}
(see~\cite{Forste:2004ie} for a recent review).  $\mathbbm{T}^6$ can be
parametrized by three complex coordinates $z_i$ ($i=1,2,3$). Then we denote
$\theta = \diag(\mathrm{e}^{2\pi\,\I\, v_1},\mathrm{e}^{2\pi\,\I\,
v_2},\mathrm{e}^{2\pi\,\I\, v_3})$. For example, in \Z6-II orbifolds one has
$v_i=(1/6,1/3,-1/2)$. A model is defined by the compactification lattice, the
twist vector $v_i$,  the shift $V$ and the Wilson lines $W_\alpha$. Given these
data, the massless spectrum (at the orbifold point) is completely determined
(for recent explicit examples see
e.g.~\cite{Kobayashi:2004ya,Buchmuller:2006ik}). A rather common feature of
these constructions is the occurrence of a so-called `anomalous \U1', 
$\U1_\mathrm{anom}$,~\cite{Casas:1987us,Kobayashi:1996pb} (cf.\
section~\ref{sec:DiscreteAnomalies}), which implies that, at one-loop, a FI
$D$-term is induced~\cite{Dine:1987xk}. As we shall see, the `anomalous' \U1
plays a prominent role in the discussion of discrete anomalies.

\subsection{Stringy discrete symmetries}
\label{sec:DiscreteOrbifoldSymmetries}

Couplings on heterotic orbifolds are governed by certain selection rules
\cite{Hamidi:1986vh,Dixon:1986qv}  (see also
\cite{Kobayashi:1991rp,Kobayashi:2004ya,Buchmuller:2006ik, Lebedev:2007hv}), 
some of which can be interpreted as discrete symmetries of the effective field
theory emerging as `low-energy' limit in these constructions. These symmetries
fall into two classes, depending on whether they reflect space group rules or
$R$-charge (or $H$-momentum) conservation.

\subsubsection{Space group rules}

The space group selection rules are stated by
\begin{equation}\label{eq:SpaceGroupRule}
 \prod_r (\theta^{k^{(r)}},n^{(r)}_\alpha\,e_\alpha)
 ~\simeq~(\mathbbm{1},0)\;,
\end{equation}
where we label the states entering the coupling by $r$.
$(\theta^{k^{(r)}},n^{(r)}_\alpha\,e_\alpha)$ is the space group
element representing the string boundary condition with 
$n^{(r)}_\alpha =$ integer, $e_\alpha$ are lattice vectors defining 
$\mathbbm{T}^6$,  and `$\simeq$' means that
the product on the l.h.s.\ lies in the same equivalence class as the identity
element. The rotational part of \eqref{eq:SpaceGroupRule} gives rise
to the point group selection rule, and here we refer to it as the
$k$-rule, which in \Z6-II orbifolds reads
\begin{equation}\label{eq:k-rule}
 \sum_r k^{(r)}~=~0 \mod 6 \;.             
\end{equation}
The translational part can be rewritten as 
\begin{subequations}\label{eq:SpaceGroupRules}
\begin{eqnarray}
 \text{\SO4\ plane}& : & 
 \sum\limits_{r=1}^n k^{(r)}\,n_2^{(r)}~=~ 0\mod2\;,\\*
 & & \sum\limits_{r=1}^n k^{(r)}\,n_2^{(r)\,\prime}~= ~0\mod2\;,\\*
 \label{eq:n2primeRule}
  \text{\SU3\ plane} & : & 
 \sum\limits_{r=1}^n k^{(r)}\, n_3^{(r)} ~=~ 0\mod3\;.
\end{eqnarray}
\end{subequations}
The quantum numbers $n_3^{(r)}$, $n_2^{(r)}$ and $n_2^{(r)\,\prime}$ specify
the localization of the states on the orbifold; we follow the conventions of
\cite{Buchmuller:2006ik}. 

The space group rules \eqref{eq:SpaceGroupRules} can be interpreted as 
$\Z2\times\Z2'\times\Z3$ flavor symmetries, denoted
$\Z2^\mathrm{flavor}\times\Z2^{\mathrm{flavor}\,\prime}\times\Z3^\mathrm{flavor}$
in what follows. Under this symmetry, each state comes with two \Z2 charges and
one \Z3 charge,
\begin{subequations}\label{eq:qnSpaceGroup}
\begin{eqnarray}
 \Z2^\mathrm{flavor} & : & q_2~=~k\,n_2 \mod 2\;,\\
 \Z2^{\mathrm{flavor}\,\prime} & : & q_2'~=~k\,n_2' \mod 2\;,\\
 \Z3^\mathrm{flavor} & : & q_3~=~k\,n_3 \mod3\;.
 \end{eqnarray}
\end{subequations}
In models where certain Wilson lines are absent, these symmetries combine with
permutation symmetries of equivalent fixed points to non-Abelian discrete flavor
symmetries~\cite{Kobayashi:2004ya,Kobayashi:2006wq}. 
As we are interested in anomalies, we focus on the Abelian subgroups of these
discrete symmetries (cf.\ section~\ref{sec:NonAbelianDiscretAnomalies}).

\subsubsection{Discrete $\boldsymbol{R}$-symmetries}

The discrete $R$-symmetries in \Z6-II orbifolds based on the Lie lattice
$\G2\times\SU3\times\SO4$ are expressed
by~\cite{Kobayashi:2004ya,Buchmuller:2006ik}
\begin{subequations}\label{eq:H-momentumRules}
\begin{eqnarray}
\sum_{r=1}^n R_1^{(r)} &=& -1\mod6\;, \\*
\sum_{r=1}^n R_2^{(r)} &=& -1\mod3\;, \\*
\sum_{r=1}^n R_3^{(r)} &=& -1\mod2\;.\label{eq:H-momentumRule3}
\end{eqnarray}
\end{subequations}
Hereby, $R_i^{(r)}$ denotes the $i^\mathrm{th}$ component of the $H$-momentum
of  the bosonic components of chiral superfields,
\begin{equation}\label{eq:DiscreteRcharge}
 R_i~=~ q_{\mathrm{sh},i}-\Delta N_i\;,
\end{equation}
where $ q_{\mathrm{sh},i}$ denote the SO(6) shifted momenta of bosonic states
and $\Delta N_i=\widetilde{N}_i-\widetilde{N}_i^*$ is the difference of
oscillator numbers  $\widetilde{N}_i,\widetilde{N}_i^*$. For twisted sectors, it can be shown that
$q_{\mathrm{sh},i} = k\, v_i - \text{int}(k\, v_i)$, with $\text{int}(k\, v_i)$ being the smallest
integer, such that $\text{int}(k\, v_i) \geq k\, v_i$. 

\subsubsection{Modular symmetries}

In orbifold constructions, $T$-duality transformations act as discrete
reparametrizations of the moduli space.  In general, there are three $T$-moduli
$T_i$ ($i=1,2,3$),  each of which corresponds to the $i$-th complex plane $z_i$.
For example, the moduli $T_1$, $T_2$ and $T_3$ in \Z6-II orbifolds
correspond to the overall sizes of $\G2$, $\SU3$ and $\SO4$ tori, respectively.
Modular symmetry is in a sense different from other symmetries, 
where moduli $T_i$ are singlets.
Under the modular symmetry, the moduli $T_i$ transform as
\begin{equation}\label{eq:T-duality}
 T_i~\rightarrow~\frac{a_i\,T_i -\I\, b_i}{\I\,c_i\,T_i+d_i} \;,
\end{equation}
where $a_i,~b_i,~c_i,~d_i \in \mathbb{Z}$ and $a\,d-b\,c =1$.
The K\"ahler potential $K_\mathrm{matter}$ of matter fields $\Phi^{(f)}$
depends in general on the $T_i$ moduli as 
\begin{equation}
 K_\mathrm{matter}~=~\prod_i \left(T_i + \overline{T_i}\right)^{m_i}\,
 |\Phi^{(f)}|^2\;,
\end{equation}
where the so-called modular weights $m_i$ are given 
by~\cite{Dixon:1989fj,Louis:1991vh,Ibanez:1992hc} 
\begin{equation}\label{eq:ModularWeights}
 m_i~=~\left\{\begin{array}{ll}
 1\;,\quad & \text{if}\: q_{\mathrm{sh},i}=-1\;,\\
 0\;,\quad & \text{if}\: q_{\mathrm{sh},i}=0\;,\\
 q_{\mathrm{sh},i}+1-\Delta N_i\;,\quad  & \text{if}\: q_{\mathrm{sh},i}\ne0,-1\;.
 \end{array}\right.
\end{equation}
We require that the K\"ahler potential $K_\mathrm{matter}$ be invariant 
under \eqref{eq:T-duality}.
This implies that the matter fields with the modular weight $m_i$ 
transform under \eqref{eq:T-duality} as the following chiral 
rotation:\footnote{The K\"ahler potential of moduli fields, 
$K_\mathrm{moduli} = - \sum_i\ln (T_i +\bar T_i)$ is not invariant 
under \eqref{eq:T-duality}.
$T$-duality invariance requires that the holomorphic 
superpotential $W$ transform as 
$W \rightarrow W\prod_i(T_i + \bar T_i)^{-1}$, such that the 
combination $G = K_\mathrm{moduli} +K_\mathrm{matter}  +\ln |W|^2$, which 
appears in the supergravity Lagrangean, is invariant.
}
\begin{equation}
 \Phi^{(f)}~\rightarrow~\Phi^{(f)}\,\prod_i 
 \left(\I\,c_i\,T_i+d_i\right)^{m_i}\;.
\end{equation}
Once the $T_i$ attain vacuum expectation values, these symmetries are (in
general) completely broken. 
That is, the $T$-duality symmetries are not
expected to contribute to discrete symmetries which survive to low energies.

\subsection{Discrete anomalies on orbifolds}

According to the various discrete symmetries described in the previous
subsections, we now define the corresponding anomaly coefficients. We further
conduct a scan over many models, based on several orbifold geometries, and
elicit whether there the symmetries of
subsection~\ref{sec:DiscreteOrbifoldSymmetries} are anomalous or not.

\subsubsection{$\boldsymbol{\Z{n}^\mathrm{flavor}}$ anomalies}

Let us start by studying anomalies in the $\Z{n}^\mathrm{flavor}$ symmetries. A
special class of $\Z{n}^\mathrm{flavor}$ anomalies is given by 
\begin{equation}\label{eq:FlavorAnomaly}
 A_{\Z{n}^\mathrm{flavor}-G-G}
 ~=~
 \frac{1}{n}\,\sum\limits_{\boldsymbol{r}^{(f)}}
 q_n^{(f)}\,2\,\ell(\boldsymbol{r}^{(f)})\;,
\end{equation}
where the sum extends over all non-trivial representations
$\boldsymbol{r}^{(f)}$ of a non-Abelian gauge factor $G$ and the $q_n^{(f)}$
are defined in \eqref{eq:qnSpaceGroup}. $A_{\Z{n}^\mathrm{flavor}-G-G}$ is
only defined up to twice the smallest non-vanishing Dynkin index $\ell_{min} =
\text{min}\big\{\ell(\boldsymbol{r}^{(f)})\big\}$ that appears, i.e.\ up to 1 if
fundamental representations of \SU{N} groups are present.

We have investigated various heterotic orbifolds, and find that, in general,
they exhibit flavor anomalies (see appendices~\ref{sec:KRZAnomalies}
and~\ref{sec:BHLRAnomalies} for specific examples). 

\subsubsection{Discrete $\boldsymbol{R}$ anomalies}

The $R$ anomalies are given by~\cite{Araki:2007ss}
\begin{equation}\label{eq:DiscreteRanomaly}
 A_{G}^{R_i}~=~ -c_2(G)
 +\sum\limits_{\boldsymbol{r}^{(f)}}\left(R_i^{(f)}+\frac{1}{2}\right)\,2\,
 \ell(\boldsymbol{r}^{(f)})\,,
\end{equation}
with $c_2$ denoting the quadratic Casimir. The sum extends over all irreps
$\boldsymbol{r}^{(f)}$ denoting the representation of the field
$f$ w.r.t.\ the gauge factor $G$. The discrete $R$ charges in this orbifold
are only defined modulo $(6,3,2)$. Therefore, the anomalies can only be
specified up to $(6,3,2)$ times twice the smallest non-vanishing Dynkin index 
$\ell_{min}$ appearing in the sum in~\eqref{eq:DiscreteRanomaly}.

We find empirically that the $R$ anomalies are not universal (for specific
examples see appendices~\ref{sec:KRZAnomalies} and~\ref{sec:BHLRAnomalies}).

\subsubsection{$\boldsymbol{T}$-duality anomalies}
\label{sec:Tanomalies}

By considering the one-loop effective supersymmetric Lagrangean, one finds that
the gauge coupling constant is not invariant under the discrete modular group of
$T$-duality transformations. The coefficients of this $T$-duality anomaly are
given by~\cite{Louis:1991vh,Derendinger:1991hq,Ibanez:1992hc,Araki:2007ss}
\begin{equation}\label{eq:DiscreteTanomaly}
 A_{G}^{T_i}~=~ 2\,c_2(G) 
 +\sum\limits_{\boldsymbol{r}^{(f)}}\left(2\, m_i^{(f)} -1\right)\,2\,
 \ell(\boldsymbol{r}^{(f)})\;,
\end{equation}
where $m_i^{(f)}$ denotes the modular weight of the state $\Phi^{(f)}$ w.r.t.\ the plane $i$
(cf.~\eqref{eq:ModularWeights}). 

As is well known, $T$-duality anomalies can be canceled in two different ways.
One part of it is removed by the Green-Schwarz mechanism whereas a second part
only disappears after considering one-loop threshold corrections to the gauge
coupling constants. Only universal anomalies, i.e.\ those $A_{G}^{T_i}$
in~\eqref{eq:DiscreteTanomaly} with fixed $i$ whose values do not depend on
$G$, can be canceled by the Green-Schwarz mechanism. In contrast,
cancellation of non-universal $T$-duality anomalies requires additionally
threshold corrections. According to~\cite{Ibanez:1992hc}, in orbifold models the
anomaly associated to the modulus $T_i$ is non-universal only if any of the
orbifold twists acts trivially on the corresponding $i^\mathrm{th}$ complex plane of the
underlying six-torus.  This means in particular, that for \Z6-II orbifolds the
anomalies of $T_2$ and $T_3$ are non-universal and therefore the associated
moduli appear in the threshold corrections. Further, since the orbifold twist
acts non-trivially on the first complex plane, the $T_1$-anomaly must be
universal to be completely canceled by the Green-Schwarz mechanism.

We have conducted a scan over $T$-anomaly in \Z6-II orbifold models, and confirm
that only the $T_1$-anomalies are universal (for our conventions for labeling
the two-tori see~\cite{Kobayashi:2004ya,Buchmuller:2006ik,Lebedev:2007hv}).
However, this does not imply that there are uncanceled $T$-anomalies in the
other tori. Rather, as we shall see in the next section, some $T$-anomalies are
inherited from what we will call the `$k$-anomaly', which can be canceled by the
Green-Schwarz mechanism.

Discrete anomalies can also be canceled by  the Green-Schwarz mechanism,
just like in the $\U1_\mathrm{anom}$ case~\cite{Banks:1991xj,Banks:1995ii}.
Under discrete transformation, the dilaton $S$  (more precisely the axion) gets
shifted according to \eqref{eq:DilatonTransformation}, 
\eqref{eq:axion-shift}.
Note that 
for the discrete transformation, the shift $\Lambda$ and $\theta$ are 
constant (cf.\ \cite{Araki:2006mw}), 
while for the anomalous $\U1_\mathrm{anom}$ the shift $\Lambda(x)$ and 
$\theta(x)$ are  $x$-dependent.
Hence, both forms of the K\"ahler potential \eqref{eq:Kdilaton-0} and 
\eqref{eq:Kdilaton} are invariant under the 
anomalous discrete transformation.
This implies that the term $\Sigma = \mathrm{e}^{-a\,S}$  has a definite 
charge under the discrete transformation.
Then, stringy non-perturbative effects induce 
terms of the form $\Phi_1\cdots \Phi_n\cdot\mathrm{e}^{-a\,S}$, 
where the $\Phi_i$
transform under (anomalous) discrete symmetries. These terms transform trivially (although
they appear to be forbidden by the discrete symmetry) because the
transformation of the fields gets compensated 
by the dilaton \cite{Banks:1991xj,Banks:1995ii}.
Note that a superpotential term $\Phi_1\cdots \Phi_n\cdot\mathrm{e}^{-a\,S}$ 
has to transform trivially both for the anomalous $\U1_\mathrm{anom}$  
and anomalous discrete symmetries.
Furthermore, anomaly cancellation by  the Green-Schwarz mechanism requires that
discrete anomalies be universal for different gauge group up to modulo the
structure \eqref{eq:deltaGS1}.
We will examine the universality conditions for discrete anomalies 
in section \ref{sec:survey}.

\subsection{Relations between discrete anomalies}

In orbifolds there are certain quantum numbers like $k$ (denoting the twisted
sector), $p_\mathrm{sh}$ (shifted $\E8\times\E8$ momentum), $q_\mathrm{sh}$
(shifted \SO8 momentum) and oscillator numbers. From these, one can derive other
useful quantum numbers such as the discrete $R$-charges and modular weights, as
defined in equations \eqref{eq:DiscreteRcharge} and \eqref{eq:ModularWeights}.
It is hence clear that the derived quantum numbers are related. On the other
hand, the discrete $R$-charges and modular weights represent discrete charges
relevant for the string selection rules. Clearly, since the $\Z{n}$ charges
derive from the same set of quantum numbers, the different $\Z{n}$ symmetries
entailing different string selection rules cannot be completely independent.

To see what this means, consider the discrete $R$-charges and the corresponding
selection rule. At first sight, one might think that the $R_1$, $R_2$ and $R_3$ 
rules in \Z6-II orbifolds entail  $\Z{36}$, $\Z{9}$ and $\Z{4}$ symmetries,
respectively. However, it is obvious that, once the $k$-rule \eqref{eq:k-rule}
is satisfied, the discrete $R$ symmetries boil down to $\Z6\times\Z3\times\Z2$
discrete symmetries. That is, one can factorize this subset of discrete
symmetries as
\begin{equation}\label{eq:FactorZN}
 \Z6^k \times [\Z6 \times \Z3 \times \Z2]_R\;.
\end{equation}

\subsubsection{A $\boldsymbol{k}$-anomaly}

This raises the question whether the $\Z6^k$ symmetry (which is implied by the
selection rule~\eqref{eq:k-rule}) has an anomaly. To clarify this, define the
$k$-anomalies as
\begin{equation}\label{eq:A_Z6-G-G}
 A_{\Z6^k-G-G}~=~\frac{1}{6}\sum_{\boldsymbol{r}^{(f)}}
 k^{(f)}\,2\,\ell(\boldsymbol{r}^{(f)})\;,
\end{equation}
where the sum extends over all non-trivial representations of $G$. Similarly
as for the flavor anomalies, the $k$-anomaly is only defined modulo twice the
smallest non-vanishing Dynkin index $\ell_\mathrm{min}$ appearing in the sum
in~\eqref{eq:A_Z6-G-G}.  Condition~\eqref{eq:A_Z_N-G-G} implies that, if
$A_{\Z6^k-G-G}$ is not integer, one has a $\Z6^k$ anomaly.

\subsubsection{$\boldsymbol{R}$- vs.\ $\boldsymbol{k}$-anomalies}
\label{subsubsec:Rvsk}

Now let us evaluate the $R_1$ anomaly, using the prescription of~\cite{Araki:2007ss}. One has 
\begin{subequations}
\begin{eqnarray}
 A_{G}^{R_1}
 & = &
 -c_2(G)+\sum_{\boldsymbol{r}^{(f)}}\left(R_1^{(f)}+\tfrac{1}{2}\right)\,2\,
 \ell(\boldsymbol{r}^{(f)})
 \nonumber\\*
 & = &
 -c_2(G)
 +\sum_{\boldsymbol{r}^{(t)}}\Big(k^{(t)}\,v_1-\underbrace{\text{int}(k^{(t)}\,v_1)}_{=1}
 -\Delta N_1^{(t)}+\tfrac{1}{2}\Big)\,2\,\ell(\boldsymbol{r}^{(t)})\nonumber\\
 & & {}
 +\sum\limits_{\boldsymbol{r}^{(u)}}\left(R_1^{(u)}+\tfrac{1}{2}\right)\,2\,
 \ell(\boldsymbol{r}^{(u)})
 \nonumber\\*
 & = &
 A_{\Z6^k-G-G}-
 c_2(G)
 -\sum_{\boldsymbol{r}^{(t)}}\left(\Delta N_1^{(t)}+\tfrac{1}{2}\right)\,2\,
 \ell(\boldsymbol{r}^{(t)})\nonumber\\
 & & {}
 +\sum\limits_{\boldsymbol{r}^{(u)}}\left(R_1^{(u)}+\tfrac{1}{2}\right)\,2\,
 \ell(\boldsymbol{r}^{(u)})
 \;,\label{eq:AR1-Ga-Ga}
\end{eqnarray}
where we have used that $v_1=1/6$. The summations $\sum_{\boldsymbol{r}^{(u)}}$ and
$\sum_{\boldsymbol{r}^{(t)}}$ extend, respectively, over untwisted
and twisted representations of the gauge factor $G$. This calculation shows that $A_{G}^{R_1}$ and
$A_{\Z6^k-G-G}$ are related. Repeating the calculation for $R_2$ and $R_3$ yields
\begin{eqnarray}
 A_{G}^{R_2}
 & = &
 2\,A_{\Z6^k-G-G}
 - c_2(G)
 +\sum_{\boldsymbol{r}^{(t)}}\left(-\Delta N_2^{(t)}+\tfrac12 - \text{int}(k^{(t)}\,v_2)\right)\,2\,
 \ell(\boldsymbol{r}^{(t)})
 \nonumber \\*
 & & {}
 +\sum\limits_{\boldsymbol{r}^{(u)}}\left(R_2^{(u)}+\tfrac12\right)\,2\,
 \ell(\boldsymbol{r}^{(u)})
 \;,\label{eq:AR2-Ga-Ga}\\
 A_{G}^{R_3}
 & = &
 3\,A_{\Z6^k-G-G} - c_2(G)
 +\sum_{\boldsymbol{r}^{(t)}}\left(-\Delta N_3^{(t)}+ \tfrac12 - \text{int}(k^{(t)}\,v_3)\right)\,2\,
 \ell(\boldsymbol{r}^{(t)})
 \nonumber \\*
 & & {}
 +\sum\limits_{\boldsymbol{r}^{(u)}}\left(R_3^{(u)}+\tfrac12\right)\,2\,
 \ell(\boldsymbol{r}^{(u)})
 \;.\label{eq:AR3-Ga-Ga}
\end{eqnarray}
\label{eq:AR-Ga-Ga}
\end{subequations}
That is, whenever $A_{\Z6^k-G-G}$ is non-zero, the $R_i$ anomalies can be fractional.

\subsubsection{An `anomalous space group element'}

In this subsection, we put the $k$- and $\Z3^\mathrm{flavor}$ anomalies into a
greater perspective. It turns out that they can be related to the so-called
`anomalous \U1' direction. Denote the corresponding generator by
$\mathsf{t}_\mathrm{anom}$.\footnote{In heterotic orbifolds, the normalization of
$\mathsf{t}_\mathrm{anom}$ is determined, so that the first equality sign in
\eqref{eq:deltaGS1} represents a non-trivial condition which can be used to check
the consistency of the model.} Obviously, $\mathsf{t}_\mathrm{anom}$ is a function
of the input, i.e.\ shift and Wilson lines, 
\begin{equation}
 \mathsf{t}_\mathrm{anom}~=~ \mathsf{t}_\mathrm{anom}(V,\{W_\alpha\})\;.
\end{equation}
This direction is fixed up to rescaling, our conventions are to
normalize $\mathsf{t}_\mathrm{anom}$ such that (for
$\mathsf{t}_\mathrm{anom}\ne0$)
\begin{equation}\label{eq:tanomNormalizationHet}
 \sum_i\frac{\mathsf{t}_\mathrm{anom}\cdot p_\mathrm{sh}^{(i)}}{\mathsf{t}_\mathrm{anom}\cdot\mathsf{t}_\mathrm{anom}}
 ~=~12\;,
\end{equation}
where the sum extends over all states.\footnote{This normalization differs from
the one used in subsection~\ref{sec:DiscreteAnomalies} above
equation~\eqref{eq:deltaGS1}. In heterotic orbifolds, one can use the scalar
product of the $\E8\times\E8$ lattice, which also appears in
\eqref{eq:tanomNormalizationHet}. With this scalar product,
$\widehat{\mathsf{t}}_\mathrm{anom}$ fulfills
$\widehat{\mathsf{t}}_\mathrm{anom}\cdot\widehat{\mathsf{t}}_\mathrm{anom}~=~1/2$.}
Together with the other properties $\U1_\mathrm{anom}$, this implies
\begin{equation}
 \mathsf{t}_\mathrm{anom}~=~\frac{1}{12}
 \sum_i p_\mathrm{sh}^{(i)}\;.
\end{equation}
Now perform a Weyl rotation of the input,
\begin{equation}\label{eq:Weyl}
 (V,\{W_\alpha\})~\to~(\Omega\,V,\{\Omega\,W_\alpha\})
\end{equation} 
with $\Omega\in\mathcal{W}$ and $\mathcal{W}$ denoting the Weyl group. This is
nothing but a change of the basis, hence
\begin{equation}
 \mathsf{t}_\mathrm{anom}~\to~\Omega\,\mathsf{t}_\mathrm{anom}
\end{equation}
under \eqref{eq:Weyl}. This fixes $\mathsf{t}_\mathrm{anom}$ to be a linear
superposition of $V$ and the $W_\alpha$ with coefficients that are invariant
under Weyl transformations. Because we are working on the lattice
$\Lambda_{\E8\times\E8}$, this relation holds only up to lattice vectors, i.e.\
\begin{equation}\label{eq:DecompositionAnomalousU(1)}
 \mathsf{t}_\mathrm{anom}~=~
 k^\mathrm{anom}\,V + \sum_\alpha n_\alpha^\mathrm{anom}\,W_\alpha +\lambda\;,
\end{equation}
where $\lambda\in\Lambda_{\E8\times\E8}$ is a lattice vector. This relation between
$\mathsf{t}_\mathrm{anom}$ and the orbifold parameters indicates that
the presence of an anomalous U(1) can be attributed to a geometrical operation in the six
dimensional compactified space. This transformation is then encoded in the space group element
$g^\mathrm{anom}=(\theta^{k^\mathrm{anom}},n_\alpha^\mathrm{anom}\,e_\alpha)$. 

We would like to comment that one cannot trade $k^\mathrm{anom}$ for
$n^\mathrm{anom}_\alpha$ (and vice versa) as long as $0\le k^\mathrm{anom}<N$
and $0\le n^\mathrm{anom}_\alpha<N_\alpha$ with $N_\alpha$ denoting the order of
the Wilson line. That is, the coefficients $k^\mathrm{anom}$ and
$n^\mathrm{anom}_\alpha$ are fixed mod $N$ and $N_\alpha$, respectively.
Further, if $\mathsf{t}_\mathrm{anom}\in\Lambda_{\E8\times\E8}$, one has
$k^\mathrm{anom}=n^\mathrm{anom}_\alpha=0$, i.e.\ if  $k^\mathrm{anom}$ or
$n^\mathrm{anom}_\alpha$ are non-zero, one can infer that
$\mathsf{t}_\mathrm{anom}\ne0$, but the converse is in general not true.

As we shall see in the next section, it turns out that the coefficients
$k^\mathrm{anom}$ and $n_\alpha^\mathrm{anom}$  are related to the $k$- and
flavor anomalies.  We have verified that the
decomposition~\eqref{eq:DecompositionAnomalousU(1)} is possible, i.e.\ that
there exist $k^\mathrm{anom}$ and $n^\mathrm{anom}_\alpha$ such that
$\left[\mathsf{t}_\mathrm{anom}-\left( k^\mathrm{anom}\,V + \sum_\alpha
n_\alpha^\mathrm{anom}\,W_\alpha\right)\right]\in\Lambda_{\E8\times\E8}$, for
several $\Z{N}$ and $\Z{N}\times \Z{M}$ orbifolds with and without Wilson
lines. 

\subsubsection{Survey of anomaly relations}
\label{sec:survey}

As we have seen, not all discrete anomalies are independent in orbifold
constructions. Specifically, we found that the $k$- and $R$ anomalies are
related by~\eqref{eq:AR-Ga-Ga}. Given the decomposition
\eqref{eq:DecompositionAnomalousU(1)}, one is tempted to suspect that discrete
anomalies are related to and determined by the coefficients $k^\mathrm{anom}$
and $n^\mathrm{anom}_\alpha$. To figure out whether this is so, we have
conducted a scan over several thousands of models with various geometries and 
have calculated the $k$-, $R$- and $T$-duality anomalies. We obtain the
following (empirical) relations: 

\paragraph{~\textbullet~Relation between the $\boldsymbol{k}$-anomaly and
$\boldsymbol{k^\mathrm{anom}}$:}
 \begin{equation}\label{eq:def_kanom}
 A_{\Z6^k-G-G}~=~\frac{k_\mathrm{anom}}{6}\mod 1\;.
 \end{equation}
In particular, the $A_{\Z6^k-G-G}$ anomalies are universal. Furthermore, the
mixed $\Z6^k-\mathrm{grav}-\mathrm{grav}$ anomaly
\begin{equation}
  A_{\Z6^k-\mathrm{grav}-\mathrm{grav}}
  ~=~
  \sum_{m} k^{(m)}\cdot \dim\boldsymbol{R}^{(m)}\;
\end{equation} 
turns out to be always $\,0~\text{mod}~3$, thus consistent with the anomaly
constraints~\eqref{eq:ZNAnomalyConstraints}.

\paragraph{~\textbullet~Relation between $\boldsymbol{A_{\Z\alpha^\mathrm{flavor}-G-G}}$ and
$\boldsymbol{n_\alpha^\mathrm{anom}}$:}
\begin{equation}
 A_{\Z3^\mathrm{flavor}-G-G}~=~\frac{n_3^\mathrm{anom}}{3}\mod 1\;.
\end{equation}

\begin{equation}
A_{\Z2^\mathrm{flavor}-G-G}~=~\frac{n_2^\mathrm{anom}}{2}\mod 1\;.
\end{equation}
These anomalies turn out to be universal for different gauge groups in 
the models under consideration.

\paragraph{~\textbullet~Relation between the $\boldsymbol{k}$- and
$\boldsymbol{R_i}$ anomalies:}
Only if there is a $k$-anomaly, the $R$ anomalies can be fractional. 
We find that the $R_i$-anomalies are `inherited' from the $k$-anomaly,
specifically
\begin{subequations}\label{eq:RkanomaliesMod}
\begin{eqnarray}
 A_{G}^{R_1} & =  & \phantom{2\, }A_{\Z6^k-G-G}\mod 1\;,\\
 A_{G}^{R_2} & =  & 2\, A_{\Z6^k-G-G}\mod 1\;,\\
 A_{G}^{R_3} & =  & 3\, A_{\Z6^k-G-G}\mod 1 \;.
\end{eqnarray}
\end{subequations}

\paragraph{~\textbullet~Relation between the $\boldsymbol{k}$- and
$\boldsymbol{T}$-duality anomalies.}
Similarly to~\eqref{eq:RkanomaliesMod}, we have found that the $T$-duality anomaly
is related to the $k$-anomaly by 
\begin{subequations}\label{eq:TkanomaliesMod}
\begin{eqnarray}
 A_{G}^{T_1} & =  & 2\, A_{\Z6^k-G-G}\mod 1\;,\\
 A_{G}^{T_2} & =  & 4\, A_{\Z6^k-G-G}\mod 1\;,\\
 A_{G}^{T_3} & =  & 6\, A_{\Z6^k-G-G}\mod 1 \;.
\end{eqnarray}
\end{subequations}
These statements apply also to the models presented in appendices~\ref{sec:KRZAnomalies}
and~\ref{sec:BHLRAnomalies}.

\paragraph{~\textbullet~Relation between the $\boldsymbol{k}$-,
$\boldsymbol{T}$-duality and $\boldsymbol{R_i}$ anomalies.}
The previous relations~\eqref{eq:RkanomaliesMod}
and~\eqref{eq:TkanomaliesMod} imply
\begin{subequations}\label{eq:RTkanomaliesMod}
\begin{eqnarray}
 A_{G}^{T_1} - A_{G}^{R_1} & = & \phantom{2\, }A_{\Z6^k-G-G}\mod 1\;,\\
 A_{G}^{T_2} - A_{G}^{R_2} & = & 2\, A_{\Z6^k-G-G}\mod 1\;,\\
 A_{G}^{T_3} - A_{G}^{R_3} & = & 3\, A_{\Z6^k-G-G}\mod 1 \;.
\end{eqnarray}
\end{subequations}

To summarize, we have conducted a search for discrete anomalies in heterotic
orbifolds. As in previous searches \cite{Banks:1991xj,Dine:2004dk},\footnote{Our
findings are not completely consistent with the relations presented in
\cite{Araki:2007ss}.} we find that all basic discrete anomalies are universal in
the models  we studied, and all anomalies can be canceled by the discrete
Green-Schwarz mechanism.
We identify previously unknown
relations between the occurrence of discrete anomalies and the so-called
`anomalous \U1'. The anomalous \U1 is in one-to-one correspondence to the
`anomalous space group element' $g^\mathrm{anom}$, whose gauge embedding is the
generator of the `anomalous' \U1. 
$T$-duality anomalies can be canceled by two ways: 
the Green-Schwarz mechanism and $T$-dependent 
threshold corrections as said in section~\ref{sec:Tanomalies}.
It is widely believed \cite{Ibanez:1992hc}
that $T$-dependent threshold corrections 
would be non-universal and there would be no certain relation 
among $T$-duality anomalies for $T_i$, which appear in 
threshold corrections, e.g.\ $T_2$ and $T_3$ in  \Z6-II orbifolds.
On the other hand, our (empirical) results \eqref{eq:TkanomaliesMod}, which have
been checked in several thousands of models with different geometries, 
show that there exist certain relations among $T$-duality anomalies.
That is, $T$-duality anomalies are related to some basic anomalies that are cancelled
only by the Green-Schwarz  mechanism.
This issue will be studied in more detail elsewhere.

\subsection{Breaking of anomalous U(1) and discrete symmetries }

As already mentioned, an `anomalous' \U1 implies the existence of a FI term,
which needs to be canceled in supersymmetric vacua (as well as in settings with
low-energy supersymmetry). That means that certain fields which have negative
$\U1_\mathrm{anom}$ charges need to attain vacuum expectation values; hence
$\U1_\mathrm{anom}$ is broken in (almost) supersymmetric vacua. In other words,
there are no `anomalous-looking' unbroken \U1 factors. The requirement of
keeping the $D$-terms of the other symmetries zero leads typically to a
situation in which more than one field attains a VEV and in which the various
VEVs are related. Achieving $D$-flatness translates in the construction of gauge
invariant monomials which carry net negative anomalous charge 
\cite{Buccella:1982nx,Font:1988tp} (see
\cite{Cleaver:1997jb,Buchmuller:2006ik,Lebedev:2007hv} for more details).

One may wonder if one could break $\U1_\mathrm{anom}$ by canceling the FI term as
usual while leaving the anomalous flavor symmetries intact. We have tried to do
this in a large set of models with `anomalous' \U1 (including the models
presented in \cite{Lebedev:2006kn}), i.e.\ we searched for gauge invariant
monomials with net negative charge under $\U1_\mathrm{anom}$ whose constituents
transform trivially under the anomalous discrete symmetries. In most models it
is hard, if not impossible, to find such a monomial. In other words, according
to what we find, the requirement of keeping supersymmetry unbroken forces one
not only to break the `anomalous' \U1, as is well known, but generically also
implies that `anomalous' discrete symmetries get broken (which is somewhat
surprising because they do of course not have a $D$-term).  However, in a couple
of models we did find a monomial whose constituents transform trivially under
some of the anomalous discrete symmetries. In these models, an
anomalous $\Z2$ subgroup of the original $\Z6^k$ remains unbroken. We posted the
details of the model at a web site \cite{WebTables:2008da}.
Implications will be studied elsewhere.

\section{Conclusions}
\label{sec:Summary}

We have studied various aspects of discrete anomalies. We started by
reproducing the well-known anomaly constraints for \Z{N} symmetries, taking a
different route than usual, namely using the path integral approach. Unlike in the
conventional approach, our derivation does not rely on contributions from heavy
Majorana fermions; only massless fermions enter the computation. We have used the
path integral approach to derive anomaly constraints for non-Abelian discrete
symmetries; the constraints are given in equation
\eqref{eq:NonAbelianAnomalyConstraints}. 

In the second part of the study, we have explored discrete anomalies in
string-derived models, focusing on heterotic orbifolds. We find that discrete
anomalies can only occur if there is an `anomalous' \U1. One can then rotate the
anomalous symmetries into two basic symmetries, corresponding to the rotational
and translational part of the space group selection rules, i.e.\ the $k$ rule
and $n_\alpha$ rules. All other anomalies, such as $R_i$-anomalies and
$T$-duality anomalies, derive from these basic anomalies. The coefficients of
the basic  anomalies are connected to an 'anomalous space group element', whose 
gauge embedding arises from the generator of the 'anomalous \U1'. We find that
the basic anomalies are always universal, such that they might be canceled by
the same Green-Schwarz mechanism that cancels the \U1 anomaly.

We have also searched for models where the `anomalous' \U1 symmetry can be
broken (i.e.\ the FI term can be canceled) without breaking the
`anomalous' discrete symmetries. While it is hard to find a model with these
properties, we could find a few examples in which an anomalous $\Z2$ symmetry
survives. The implication of these anomalous $\Z2$ symmetries will be discussed
elsewhere.

Of course, discrete and continuous symmetries that are broken by a suppressed
VEV, as is the case in the `anomalous' \U1, are known to be a useful tool in
model building. Indeed, our results indicate that in string models discrete
cousins of `anomalous' \U1 symmetries are frequently present, whereby, according
to what we find, cancellation of the FI term triggers symmetry
breakdown. Since the FI term is loop suppressed, the vacuum
expectation value of the field that breaks the symmetry can be small. The
emerging approximate symmetries can play an important role in understanding the
observed pattern of fermion masses and mixings.

\subsubsection*{Acknowledgments}

We acknowledge discussions with K.~Fujikawa, R.N.~Mohapatra, H.P.~Nilles and
S.~Raby. 
We would like to thank the Summer Institute 2007 (held at Fuji-Yoshida), where
this work was initiated, and the Aspen Center for Physics, where some of the
work has been carried out, for hospitality and support. 
This research was
supported by the Grand-in-Aid for 
Scientific Research No.~20540266 and No.~18540257 from the 
Ministry of Education, Culture, Sports, Science and Technology of
Japan, 
the DFG cluster of excellence Origin and Structure of the Universe,
the European Union 6th framework program MRTN-CT-2004-503069 "Quest for
unification", MRTN-CT-2004-005104 "ForcesUniverse", MRTN-CT-2006-035863
"UniverseNet" and SFB-Transregios 27 "Neutrinos and Beyond" and 33 "The Dark
Universe" by Deutsche Forschungsgemeinschaft (DFG).

\appendix
\section{Anomalies of discrete non-Abelian $\boldsymbol{D_4}$ symmetry}
\label{app:D4}

In this appendix, we discuss anomalies of the discrete symmetry $D_4$.
The $D_4$ symmetry is one of the simplest non-Abelian discrete
symmetries.\footnote{%
The $D_4$ flavor symmetry happens to occur in certain, potentially realistic
string models \cite{Kobayashi:2004ya,Buchmuller:2006ik,
Lebedev:2006kn,Lebedev:2007hv},  which have been constructed recently  within
the framework of heterotic orbifolds.}

The non-Abelian finite group $D_4$ has eight elements, 
which can be written as products of the two generators $g$ and $h$, 
i.e.
\begin{equation}
 \mathcal{G}_{D_4} 
 ~=~ 
 \{ \mathbbm{1}, g, h, g\,h, h\,g, h\,g\,h, g\,h\,g, g\,h\,g\,h  \}\; .
\end{equation}
$D_4$ has five irreps: $\boldsymbol{2}$, $\boldsymbol{1}_{++}$, 
$\boldsymbol{1}_{+-}$, 
$\boldsymbol{1}_{-+}$ and $\boldsymbol{1}_{--}$. 
The action of $g$ and $h$ on these irreps is 
\begin{eqnarray}
\begin{array}{llll}
  \boldsymbol{2} & : & g ~=~
\begin{pmatrix}
1 & 0 \\ 0 & -1
\end{pmatrix}
, & h~ =~  
\begin{pmatrix}
0 & 1 \\ 1 & 0 
\end{pmatrix},
\\
  \boldsymbol{1}_{++} & :  & g ~=~ 1\; ,      &     h ~=~   1\;,  \\
  \boldsymbol{1}_{+-} & :  & g ~=~ 1\; ,      &     h ~=~  - 1\;,  \\
  \boldsymbol{1}_{-+} & :  & g ~=~ -1\; ,      &     h ~=~   1\;,  \\
  \boldsymbol{1}_{--} & :  & g ~=~ -1\; ,      &     h ~=~  - 1\;.  
\end{array}
\end{eqnarray}
According to our discussion in 
section~\ref{sec:NonAbelianDiscretAnomalies}, all we need to do 
for $D_4$ anomalies is to study the anomalies for the group 
elements $g$ and $h$ (or another combination).

The $D_4$ flavor symmetry can appear from  $\mathbbm{Z}_6$-II orbifold models
\cite{Kobayashi:2004ya,Kobayashi:2006wq} (and other orbifold models whose
compact spaces include the 1D $\Z2$ sub-orbifold).
In $\mathbbm{Z}_6$-II orbifold models, the group 
element $g$ corresponds to $\Z2^\mathrm{flavor}$ or 
$\Z2^{\mathrm{flavor}\,\prime}$.
There are two fixed points on the 1D $\Z2$ sub-orbifold.
Massless spectra on these two fixed points are degenerate, 
when there is no Wilson line on the 1D $\Z2$ sub-orbifold.
Then, these modes correspond to $\boldsymbol{2}$ and the group 
element $h$ corresponds to the permutation of these modes.
In $\mathbbm{Z}_6$-II orbifold models, 
only the doublet ${\bf 2}$ and the trivial singlet 
${\bf 1}_{++}$ can appear as fundamental modes.
In this case, anomalies are constrained.
We denote 
\begin{equation}
h' = 
\begin{pmatrix}
0 & -1 \\
1 & 0  \\
\end{pmatrix}
\;.
\end{equation}
Now note that $h=h'\,g$ for the doublet $\boldsymbol{2}$ and  $\det h' =1$.
Thus, all eight elements of the $D_4$ group can be written 
as products of $g$ and $h'$, and the generator $h'$ 
does not lead to anomalies.
That implies that all of $D_4$ anomalies originate from 
$\Z2^\mathrm{flavor}$ anomalies, that is, 
$D_4$ anomalies, e.g.\ anomalies for the permutation $h$, 
appear in $\mathbbm{Z}_6$-II orbifold models only if
there are $\Z2^\mathrm{flavor}$ anomalies, i.e. 
anomalies for the group element $g$.
The situation is the same for $D_4$ anomalies 
in heterotic orbifold models with the 1D $\Z2$ sub-orbifold 
such as $\Z2 \times \Z{M}$.

The situation would change if we had heterotic orbifold models  including
non-trivial singlets of the $D_4$ flavor symmetry,  in $\boldsymbol{1}_{+-}$ and
$\boldsymbol{1}_{-+}$,  because in these representation the determinants of $g$
and $h$ differ.
Indeed, in heterotic orbifold models including  the 2D $\Z4$ sub-orbifold,
non-trivial singlets can appear as fundamental modes~\cite{Kobayashi:2006wq}.
However, massless states corresponding to $\boldsymbol{1}_{+-}$ and
$\boldsymbol{1}_{-+}$ are always degenerate. This can only be changed  by
introducing a Wilson line,
which, however, breaks the $D_4$ flavor symmetry.
Thus, in these models,  non-trivial singlets  $\boldsymbol{1}_{+-}$ and
$\boldsymbol{1}_{-+}$ do not contribute to  anomalies.
Therefore, the situation is the same as  $\mathbbm{Z}_6$-II orbifold models,
that is,  all of $D_4$ anomalies originate from  $\Z2^\mathrm{flavor}$ anomalies.

\section{Sample calculation of discrete anomalies}
\label{app:GrimusLavoura}

In this appendix we present a sample calculation in order to demonstrate how the
anomaly constraints can be applied.
We will base the calculations on the Grimus-Lavoura model~\cite{Grimus:2003kq},
which is not supersymmetric.
The lepton and Higgs fields are assigned the transformation properties displayed
in table~\ref{tab:GrimusLavoura2}.
\begin{table}[h]
\begin{center}
\begin{tabular}{|c||c|c|c|c|c|c|c|}\hline
 &
 $D_e$ &
 $(D_{\mu},D_{\tau})$ &
 $e_\mathrm{R}\ \nu_{e\mathrm{R}}$ &
 $(\mu_\mathrm{R} , \tau_\mathrm{R})\ (\nu_{\mu\mathrm{R}} , \nu_{\tau\mathrm{R}})$ & 
 $\phi_1,~\phi_2$ &
 $\phi_3$ &
 $(\chi_1 , \chi_2)$ \\ \hline\hline
 $D_4$ &
 $\boldsymbol{1}_{++}$ & $\boldsymbol{2}$ & 
 	$\boldsymbol{1}_{++}$ & $\boldsymbol{2}$ & $\boldsymbol{1}_{++}$ & 
 $\boldsymbol{1}_{+-}$ & $\boldsymbol{2}$  \\ \hline
 $\SU2_\mathrm{L}$ &
 $\boldsymbol{2}$ & $\boldsymbol{2}$ & $\boldsymbol{1}$ & $\boldsymbol{1}$ & $\boldsymbol{2}$ & $\boldsymbol{2}$ & $\boldsymbol{1}$ \\ \hline
\end{tabular}
\end{center}
\caption{Transformation properties of the lepton and Higgs fields in \cite{Grimus:2003kq}.}
\label{tab:GrimusLavoura2}
\end{table}
$\phi_{2,3}$ are extra $\SU2_\mathrm{L}$ doublet Higgs fields and $\chi_{1,2}$
are extra gauge singlet Higgs fields. All quark fields are assumed to be
trivial $D_4$ singlets, i.e.\ to transform as $\boldsymbol{1}_{++}$.

Let us now calculate the anomaly coefficients of the mixed anomaly $D_4 -
\SU2_\mathrm{L} - \SU2_\mathrm{L}$.
According to our discussion in section \ref{sec:NonAbelianDiscretAnomalies}, all
we need to do is to study the generators, i.e.\ the group elements  $g$ and $h$,
in order to check whether this model is anomalous or not. As $g^2 =
h^2=\mathbbm{1}$, this then amounts to checking the conditions for $\Z2$ anomalies.
For $g$ and $h$, only $(D_{\mu},D_{\tau})$ contributes to the
calculation of the anomaly.
Hence we find
\begin{eqnarray}
\Z2^g-\SU2_\mathrm{L} - \SU2_\mathrm{L}
 & :&
 \sum_{(\boldsymbol{r}^{(f)},\boldsymbol{d}^{(f)})} 
 \frac{2\ln\det g(\boldsymbol{d}^{(f)})}{2\pi\,\I}\, 
 \ell(\boldsymbol{r}^{(f)})
 ~=~\frac{1}{2}\mod 1\;,\\
 \Z2^h-\SU2_\mathrm{L} - \SU2_\mathrm{L}
 & : &
 \sum_{(\boldsymbol{r}^{(f)},\boldsymbol{d}^{(f)})} 
 \frac{2\ln\det h(\boldsymbol{d}^{(f)})}{2\pi\,\I}\
 \ell(\boldsymbol{r}^{(f)})
 ~=~\frac{1}{2}\mod1\;.
\end{eqnarray}
Therefore, the symmetry generated by $g$ and, hence, the $D_4$ symmetry 
of this model is anomalous. 

Repeating the calculation for $\U1_Y$ yields
\begin{eqnarray}
\Z2^g-\U1_Y -\U1_Y
&  : &
\sum_{(\boldsymbol{r}^{(f)},\boldsymbol{d}^{(f)})} 
 \frac{2\ln\det g(\boldsymbol{d}^{(f)})}{2\pi\,\I} 
 \left(\frac{q_Y^{(f)}}{2}\right)^2
 ~=~\frac{1}{2}\mod 1\;,\\
 \Z2^h-\U1_Y - \U1_Y
 &  : &
 \sum_{(\boldsymbol{r}^{(f)},\boldsymbol{d}^{(f)})} 
 \frac{2\ln\det h(\boldsymbol{d}^{(f)})}{2\pi\,\I} 
 \left(\frac{q_Y^{(f)}}{2}\right)^2
 ~=~\frac{1}{2}\mod1\;,
\end{eqnarray}
where the summation runs over all non-trivial $D_4$ representations with
non-zero hypercharge.
We close by stating that the anomalies do not necessarily invalidate the model.
As discussed in the conclusions, it just means that the symmetry gets broken by
certain fields attaining VEVs, which can be suppressed.

\section{Anomalies in the KRZ model}
\label{sec:KRZAnomalies}

This appendix summarizes the discrete anomalies in the KRZ model A1
\cite{Kobayashi:2004ya}. 

\subsection{$\boldsymbol{R}$ anomalies}

We obtain for the $R$ anomalies
\begin{subequations}
\begin{eqnarray}
 A_{\SU4}^{\vec R} & = & \left(13/3,5/3,1\right)\;,\\
 A_{\SU2_\mathrm{L}}^{\vec R} & = & \left(13/3,8/3,1\right)\;,\\
 A_{\SU2_\mathrm{R}}^{\vec R} & = & \left(13/3,5/3,1\right)\;.
\end{eqnarray}
\end{subequations}
The anomalies are only fixed up to $(6,3,2)$. Here, the
$R_3$ anomalies match while the others do not. They satisfy only
\begin{equation}
  A_{\SU4}^{R_{2}}~=~A_{\SU2_\mathrm{R}}^{R_{2}}
  ~\ne~A_{\SU2_\mathrm{L}}^{R_{2}}\mod 3\;.
\end{equation}

One can repeat the analysis for the non-Abelian subgroups of the second \E8.
This leads again to the result that anomalies are not universal. The R anomalies
for the KRZ model are summarized in table~\ref{tab:RanomaliesKRZ}.

\begin{table}[h]
\begin{center}
\begin{tabular}{|l|l|l|l|}
\hline
$G$ & $R_1$ & $R_2$ & $R_3$\\
\hline
SU(4)
& $\frac{13}{3}\mod 6$ & $\frac{5}{3}\mod 3$ & $1\mod 2$\\
SU(2)$_\mathrm{L} $
& $\frac{13}{3}\mod 6$ & $\frac{8}{3}\mod 3$ & $1\mod 2$\\
SU(2)$_\mathrm{R} $
& $\frac{13}{3}\mod 6$ & $\frac{5}{3}\mod 3$ & $1\mod 2$\\
SO(10) 
& $\frac{7}{3}\mod 12$ & $\frac{17}{3}\mod 6$ & $3\mod 4$\\
SU(2)$'$ 
& $\frac{7}{3}\mod 6$ & $\frac{2}{3}\mod 3$ & $1\mod 2$\\
\hline
\end{tabular}
\end{center}
\caption{Summary of $R$ anomalies in the KRZ model.}
\label{tab:RanomaliesKRZ}
\end{table}

\subsection{Flavor anomalies in the KRZ model}
\label{sec:FlavorKRZ}

Let us calculate the flavor anomalies in the KRZ model. The \Z3 symmetry is
anomalous, but the $G-G-\Z3$ anomalies are universal (see table
\ref{tab:FlavorAnomaliesKRZ}). Note, however, that there is no gravitational \Z3
anomaly if one considers the charged fields only. This means that there is an
uncharged (modulus) field that contributes to the gravitational anomaly.

\begin{table}[h]
\begin{center}
\begin{tabular}{|l|l|l|l|}
\hline
$G$ & $\Z2$ & $\Z2'$ & $\Z3$\\
\hline
$\text{SU}(4)$ & $0\mod 1$ & $0\mod 1$ &   $\frac{1}{3}\mod 1$\\
$\text{SU}(2)_\mathrm{L}$ & $0\mod 1$ & $0\mod 1$ &   $\frac{1}{3}\mod 1$\\
$\text{SU}(2)_\mathrm{R}$ & $0\mod 1$ & $0\mod 1$ &   $\frac{1}{3}\mod 1$\\
$\SO{10}$ & $0\mod 2$ & $0\mod 2$ &   $\frac{4}{3}\mod 2$\\
$\text{SU}(2)'$ & $0\mod 1$ & $0\mod 1$ &   $\frac{1}{3}\mod 1$\\
\hline
\end{tabular}
\end{center}
\caption{Summary of \Z{n} anomalies in the KRZ model.}
\label{tab:FlavorAnomaliesKRZ}
\end{table}

\subsection{$\boldsymbol{T}$-duality anomalies}

The T-duality anomalies are calculated according to equation (38) of
\cite{Araki:2007ss}; the result is listed in table
\ref{tab:TdualityAnomaliesKRZ}.

\begin{table}[h]
\begin{center}
\begin{tabular}{|l|l|l|l|l|}
\hline
$\SU4$ & $\SU2_\mathrm{L}$ & $\SU2_\mathrm{R}$ & $\SO{10}$ & $\SU2'$ \\
\hline
$\left(\frac{62}{3},-\frac{14}{3},-2\right)$ & 
$\left(\frac{62}{3},-\frac{14}{3},-2\right)$ & 
$\left(\frac{62}{3},-\frac{14}{3},-2\right)$ & 
$\left(\frac{62}{3},\frac{34}{3},-6\right)$ & 
$\left(\frac{62}{3},-\frac{14}{3},-18\right)$ \\
\hline
\end{tabular}
\end{center}
\caption{Summary of $T$-duality anomalies in the KRZ model.}
\label{tab:TdualityAnomaliesKRZ}
\end{table}

\subsection{Anomalous U(1)}

The coefficients of the anomalous \U1 (cf.\ equation
\eqref{eq:DecompositionAnomalousU(1)}) are 
$(k^\mathrm{anom},n_2^\mathrm{anom},n_3^\mathrm{anom})=(2,0,1)$.

\section{Calculation of anomalies in the BHLR model}
\label{sec:BHLRAnomalies}

\subsection{$\boldsymbol{R}$ anomalies}

Let us now consider the model described in \cite{Buchmuller:2006ik}. 
Let us focus on the non-Abelian subgroups of the first \E8 factor, i.e.\ \SU3
and \SU2. Start with \SU3. We have 10 $\boldsymbol{3}$-plets and 10
$\overline{\boldsymbol{3}}$-plets under \SU3 (quark doublets give rise to two
$\boldsymbol{3}$-plets each). 
By performing the sum~\eqref{eq:DiscreteRanomaly}, one obtains
\begin{equation}
 A_{\SU3}^{R_i}~=~(0,1,1)\mod(6,3,2)\;.
\end{equation}

Continue with \SU2. We have 30 $\boldsymbol{2}$-plets.
By performing the sum \eqref{eq:DiscreteRanomaly},
one obtains
\begin{equation}
 A_{\SU2}^{R_i}~=~(0,0,1)\mod(6,3,2)\;.
\end{equation}

While $A_{\SU3}^{R_1}=A_{\SU2}^{R_1}\mod 6$ and
$A_{\SU3}^{R_3}=A_{\SU2}^{R_3}\mod 2$, one finds
\begin{equation}
 A_{\SU3}^{R_2}~\ne~A_{\SU2}^{R_2}\mod 3\;.
\end{equation}

\begin{table}[h]
\begin{center}
\begin{tabular}{|l|l|l|l|}
\hline
$G$ & $R_1$ & $R_2$ & $R_3$\\
\hline
$\text{SU}(3)$ & $0\mod 6$ & $1\mod 3$ & $1\mod 2$ \\
$\text{SU}(2)$ & $0\mod 6$ & $0\mod 3$ & $1\mod 2$ \\
$\text{SU}(4)$ & $3\mod 6$ & $0\mod 3$ & $1\mod 2$ \\
$\text{SU}(2)'$ & $0\mod 6$ & $1\mod 3$ & $1\mod 2$ \\
\hline
\end{tabular}
\end{center}
\caption{Summary of $R$ anomalies in the BHLR model.}
\label{tab:RanomaliesBHLR}
\end{table}

\subsection{Anomalies of discrete flavor symmetries}

The flavor anomalies (cf.\ equation \eqref{eq:FlavorAnomaly}) in this model are 
\begin{subequations}
\begin{eqnarray}
A_{\SU3}^{(\Z2,\Z2',\Z3)}& = & 
  \left\{0,0,\frac{2}{3}\right\}\;,\\
A_{\SU2}^{(\Z2,\Z2',\Z3)}& = &   
   \left\{0,0,\frac{2}{3}\right\}\;.
\end{eqnarray}
\end{subequations}
That is, the \Z3 symmetry has anomalies, but they appear to be universal. This
applies also to $\Z3-G-G$ anomalies where $G$ denotes a subgroup of the
second \E8 (see table \ref{tab:FlavorAnomaliesBHLR}).
Notice, on the other hand, that the gravitational \Z3 anomalies seem to vanish.

\subsection{$\boldsymbol{T}$-duality anomalies}

The $T$-duality anomalies are calculated according to equation (38) of
\cite{Araki:2007ss}; the result is listed in table
\ref{tab:TdualityAnomaliesBHLR}.

\subsection{Anomalous U(1)}

The coefficients of the anomalous \U1 (cf.\ equation
\eqref{eq:DecompositionAnomalousU(1)}) are 
$(k^\mathrm{anom},n_2^\mathrm{anom},n_3^\mathrm{anom})=(0,0,2)$.

\begin{table}[h]
\begin{center}
\begin{tabular}{|l|l|l|l|}
\hline
$G$ & $\Z2$ & $\Z2'$ & $\Z3$\\
\hline
$\text{SU}(3)$ & $0\mod 1$ & $0\mod 1$ &   $\frac{2}{3}\mod 1$\\
$\text{SU}(2)$ & $0\mod 1$ & $0\mod 1$ &   $\frac{2}{3}\mod 1$\\
$\text{SU}(4)$ & $0\mod 1$ & $0\mod 1$ &   $\frac{2}{3}\mod 1$\\
$\text{SU}(2)'$ & $0\mod 1$ & $0\mod 1$ &   $\frac{2}{3}\mod 1$\\
\hline
\end{tabular}
\end{center}
\caption{Summary of \Z{n} anomalies in the BHLR model.}
\label{tab:FlavorAnomaliesBHLR}
\end{table}

\begin{table}[h]
\begin{center}
\begin{tabular}{|l|l|l|l|}
\hline
$\SU3$ & $\SU2$ & $\SU4$ &  $\SU2'$ \\
\hline
 $(10,10,-6)$ & $(10,10,-6)$ & $(10,10,-6)$ & $(10,2,-2)$ \\
\hline
\end{tabular}
\end{center}
\caption{Summary of $T$-duality anomalies in the BHLR model.}
\label{tab:TdualityAnomaliesBHLR}
\end{table}

% \bibliography{Orbifold}

\begin{thebibliography}{10}

\bibitem{Krauss:1988zc}
L.~M. Krauss and F.~Wilczek, Phys. Rev. Lett. \textbf{62} (1989), 1221.
%%CITATION = PRLTA,62,1221;%%

\bibitem{Ibanez:1991hv}
L.~E. Ib{\'a}{\~n}ez and G.~G. Ross, Phys. Lett. \textbf{B260} (1991),
  291--295.
%%CITATION = PHLTA,B260,291;%%

\bibitem{Banks:1991xj}
T.~Banks and M.~Dine, Phys. Rev. \textbf{D45} (1992), 1424--1427,
  [hep-th/9109045].
%%CITATION = HEP-TH/9109045;%%

\bibitem{Frampton:1994rk}
P.~H. Frampton and T.~W. Kephart, Int. J. Mod. Phys. \textbf{A10} (1995),
  4689--4704,  [hep-ph/9409330].
%%CITATION = HEP-PH/9409330;%%

\bibitem{Fujikawa:1979ay}
K.~Fujikawa, Phys. Rev. Lett. \textbf{42} (1979), 1195.
%%CITATION = PRLTA,42,1195;%%

\bibitem{Fujikawa:1980eg}
K.~Fujikawa, Phys. Rev. \textbf{D21} (1980), 2848.
%%CITATION = PHRVA,D21,2848;%%

\bibitem{Araki:2006mw}
T.~Araki, Prog. Theor. Phys. \textbf{117} (2007), 1119--1138,
  [hep-ph/0612306].
%%CITATION = HEP-PH/0612306;%%

\bibitem{Dine:2004dk}
M.~Dine and M.~Graesser, JHEP \textbf{01} (2005), 038,  [hep-th/0409209].
%%CITATION = HEP-TH/0409209;%%

\bibitem{Bertlmann:1996xk}
R.~A. Bertlmann, Oxford, UK: Clarendon (1996) 566 p. (International series of
  monographs on physics: 91).

\bibitem{AlvarezGaume:1983ig}
L.~Alvarez-Gaume and E.~Witten, Nucl. Phys. \textbf{B234} (1984), 269.
%%CITATION = NUPHA,B234,269;%%

\bibitem{AlvarezGaume:1984dr}
L.~Alvarez-Gaume and P.~H. Ginsparg, Ann. Phys. \textbf{161} (1985), 423.
%%CITATION = APNYA,161,423;%%

\bibitem{Fujikawa:1986hk}
K.~Fujikawa, S.~Ojima, and S.~Yajima, Phys. Rev. \textbf{D34} (1986), 3223.
%%CITATION = PHRVA,D34,3223;%%

\bibitem{Fujikawa:2004cx}
K.~Fujikawa and H.~Suzuki, Oxford, UK: Clarendon (2004) 284 p.

\bibitem{Rohlin:1959}
V.~Rohlin, Dokl. Akad. Nauk. \textbf{128} (1959), 980 --983.

\bibitem{Csaki:1997aw}
C.~Csaki and H.~Murayama, Nucl. Phys. \textbf{B515} (1998), 114--162,
  [hep-th/9710105].
%%CITATION = HEP-TH/9710105;%%

\bibitem{Ibanez:1991pr}
L.~E. Ib{\'a}{\~n}ez and G.~G. Ross, Nucl. Phys. \textbf{B368} (1992), 3--37.
%%CITATION = NUPHA,B368,3;%%

\bibitem{Ibanez:1992ji}
L.~E. Ib{\'a}{\~n}ez, Nucl. Phys. \textbf{B398} (1993), 301--318,
  [hep-ph/9210211].
%%CITATION = HEP-PH/9210211;%%

\bibitem{Babu:2002tx}
K.~S. Babu, I.~Gogoladze, and K.~Wang, Nucl. Phys. \textbf{B660} (2003),
  322--342,  [hep-ph/0212245].
%%CITATION = HEP-PH/0212245;%%

\bibitem{Dreiner:2005rd}
H.~K. Dreiner, C.~Luhn, and M.~Thormeier, Phys. Rev. \textbf{D73} (2006),
  075007,  [hep-ph/0512163].
%%CITATION = HEP-PH 0512163;%%

\bibitem{Green:1984sg}
M.~B. Green and J.~H. Schwarz, Phys. Lett. \textbf{B149} (1984), 117--122.
%%CITATION = PHLTA,B149,117;%%

\bibitem{Kobayashi:1996pb}
T.~Kobayashi and H.~Nakano, Nucl. Phys. \textbf{B496} (1997), 103--131,
  [hep-th/9612066].
%%CITATION = HEP-TH 9612066;%%

\bibitem{Banks:1995ii}
T.~Banks and M.~Dine, Phys. Rev. \textbf{D53} (1996), 5790--5798,
  [hep-th/9508071].
%%CITATION = HEP-TH/9508071;%%

\bibitem{ArkaniHamed:1998nu}
N.~Arkani-Hamed, M.~Dine, and S.~P. Martin, Phys. Lett. \textbf{B431} (1998),
  329--338,  [hep-ph/9803432].
%%CITATION = HEP-PH/9803432;%%

\bibitem{Leurer:1993gy}
M.~Leurer, Y.~Nir, and N.~Seiberg, Nucl. Phys. \textbf{B420} (1994), 468--504,
  [hep-ph/9310320].
%%CITATION = HEP-PH/9310320;%%

\bibitem{Binetruy:1996xk}
P.~Bin{\'e}truy, S.~Lavignac, and P.~Ramond, Nucl. Phys. \textbf{B477} (1996),
  353--377,  [hep-ph/9601243].
%%CITATION = HEP-PH/9601243;%%

\bibitem{Dixon:1985jw}
L.~J. Dixon, J.~A. Harvey, C.~Vafa, and E.~Witten, Nucl. Phys. \textbf{B261}
  (1985), 678--686.
%%CITATION = NUPHA,B261,678;%%

\bibitem{Dixon:1986jc}
L.~J. Dixon, J.~A. Harvey, C.~Vafa, and E.~Witten, Nucl. Phys. \textbf{B274}
  (1986), 285--314.
%%CITATION = NUPHA,B274,285;%%

\bibitem{Forste:2004ie}
S.~F{\"o}rste, H.~P. Nilles, P.~K.~S. Vaudrevange, and A.~Wingerter, Phys. Rev.
  \textbf{D70} (2004), 106008,  [hep-th/0406208].
%%CITATION = HEP-TH 0406208;%%

\bibitem{Kobayashi:2004ya}
T.~Kobayashi, S.~Raby, and R.-J. Zhang, Nucl. Phys. \textbf{B704} (2005),
  3--55,  [hep-ph/0409098].
%%CITATION = HEP-PH 0409098;%%

\bibitem{Buchmuller:2006ik}
W.~Buchm{\"u}ller, K.~Hamaguchi, O.~Lebedev, and M.~Ratz, Nucl. Phys.
  \textbf{B785} (2007), 149--209,  [hep-th/0606187].
%%CITATION = HEP-TH 0606187;%%

%\bibitem{Casas:1988se}
%J.~A. Casas and C.~Mu{\~n}oz, Phys. Lett. \textbf{B209} (1988), 214.
%%CITATION = PHLTA,B209,214;%%


\bibitem{Casas:1987us}
J.~A.~Casas, E.~K.~Katehou and C.~Mu{\~n}oz, Nucl.\ Phys.\  B {\bf 317} (1989) 171.
%%CITATION = NUPHA,B317,171;%%


\bibitem{Dine:1987xk}
M.~Dine, N.~Seiberg, and E.~Witten, Nucl. Phys. \textbf{B289} (1987), 589.
%%CITATION = NUPHA,B289,589;%%

\bibitem{Hamidi:1986vh}
S.~Hamidi and C.~Vafa, Nucl. Phys. \textbf{B279} (1987), 465.
%%CITATION = NUPHA,B279,465;%%

\bibitem{Dixon:1986qv}
L.~J. Dixon, D.~Friedan, E.~J. Martinec, and S.~H. Shenker, Nucl. Phys.
  \textbf{B282} (1987), 13--73.
%%CITATION = NUPHA,B282,13;%%

\bibitem{Kobayashi:1991rp}
T.~Kobayashi and N.~Ohtsubo, Int. J. Mod. Phys. \textbf{A9} (1994), 87--126.
%%CITATION = IMPAE,A9,87;%%

\bibitem{Lebedev:2007hv}
O.~Lebedev, H.~P. Nilles, S.~Raby, S.~Ramos-S{\'a}nchez, M.~Ratz, P.~K.~S.
  Vaudrevange, and A.~Wingerter, Phys. Rev. \textbf{D77} (2007), 046013,
  [arXiv:0708.2691 [hep-th]].
%%CITATION = ARXIV:0708.2691;%%

\bibitem{Kobayashi:2006wq}
T.~Kobayashi, H.~P. Nilles, F.~Pl{\"o}ger, S.~Raby, and M.~Ratz, Nucl. Phys.
  \textbf{B768} (2007), 135--156,  [hep-ph/0611020].
%%CITATION = HEP-PH/0611020;%%

\bibitem{Araki:2007ss}
  T.~Araki, K.~S.~Choi, T.~Kobayashi, J.~Kubo and H.~Ohki,
  %``Discrete R-symmetry anomalies in heterotic orbifold models,''
  Phys.\ Rev.\  D {\bf 76}, 066006 (2007)
  [arXiv:0705.3075 [hep-ph]].
  %%CITATION = PHRVA,D76,066006;%%

\bibitem{Dixon:1989fj}
L.~J. Dixon, V.~Kaplunovsky, and J.~Louis, Nucl. Phys. \textbf{B329} (1990),
  27--82.
%%CITATION = NUPHA,B329,27;%%

\bibitem{Louis:1991vh}
J.~Louis, Talk presented at the 2nd International Symposium on Particles,
  Strings and Cosmology, Boston, MA, Mar 25-30, 1991.

\bibitem{Ibanez:1992hc}
L.~E. Ib{\'a}{\~n}ez and D.~L{\"u}st, Nucl. Phys. \textbf{B382} (1992),
  305--364,  [hep-th/9202046].
%%CITATION = HEP-TH/9202046;%%

\bibitem{Derendinger:1991hq}
J.~P. Derendinger, S.~Ferrara, C.~Kounnas, and F.~Zwirner, Nucl. Phys.
  \textbf{B372} (1992), 145--188.
%%CITATION = NUPHA,B372,145;%%

\bibitem{Buccella:1982nx}
F.~Buccella, J.~P. Derendinger, S.~Ferrara, and C.~A. Savoy, Phys. Lett.
  \textbf{B115} (1982), 375.
%%CITATION = PHLTA,B115,375;%%

\bibitem{Font:1988tp}
A.~Font, L.~E. Ib{\'a}{\~n}ez, H.~P. Nilles, and F.~Quevedo, Nucl. Phys.
  \textbf{B307} (1988), 109, Erratum {\em ibid.} {\bf B310}.
%%CITATION = NUPHA,B307,109;%%

\bibitem{Cleaver:1997jb}
G.~Cleaver, M.~Cveti{\u{c}}, J.~R. Espinosa, L.~L. Everett, and P.~Langacker,
  Nucl. Phys. \textbf{B525} (1998), 3--26,  [hep-th/9711178].
%%CITATION = HEP-TH 9711178;%%

\bibitem{Lebedev:2006kn}
O.~Lebedev, H.~P. Nilles, S.~Raby, S.~Ramos-S{\'a}nchez, M.~Ratz, P.~K.~S.
  Vaudrevange, and A.~Wingerter, Phys. Lett. \textbf{B645} (2007), 88,
  [hep-th/0611095].
%%CITATION = HEP-TH 0611095;%%

\bibitem{WebTables:2008da}
T.~Araki, T.~Kobayashi, J.~Kubo, S.~Ramos-S{\'a}nchez, M.~Ratz, and P.~K.
  Vaudrevange, \emph{Additional material for discrete anomalies}, 2008,
  {\texttt{www.th.physik.uni-bonn.de/nilles/Z6IIorbifold/anomalies/}}.

\bibitem{Grimus:2003kq}
W.~Grimus and L.~Lavoura, Phys. Lett. \textbf{B572} (2003), 189--195,
  [hep-ph/0305046].
%%CITATION = HEP-PH/0305046;%%

\end{thebibliography}
% \bibliographystyle{ArXiv}

\providecommand{\bysame}{\leavevmode\hbox to3em{\hrulefill}\thinspace}

\end{document}